\documentclass{aa}  

\usepackage{graphicx}
\usepackage{txfonts}
\usepackage{comment}
\usepackage{caption}
\usepackage{subfig}
\usepackage{lipsum}

\usepackage{tabularx}
\usepackage{booktabs}

\newcommand{\norm}[1]{\lVert#1\rVert}
\newcommand{\abs}[1]{\lvert#1\rvert}

\newcommand{\tauext}{\hat{\tau}^{\text{sph}}_{\text{ext}}}
\newcommand{\ndust}{n_\text{dust}}
\newcommand{\taumis}{\tau_{\text{abs}}^{\text{mis}}}
\newcommand{\rsph}{r_\text{sph}}
\newcommand{\tausmax}{\hat{\tau}_\text{S}^{\text{max}}}
\newcommand{\tautmax}{\hat{\tau}_\text{T}^{\text{max}}}
\newcommand{\deltaemax}{\Delta E_{\text{T}}^{\uparrow}}
\newcommand{\rla}{r_\text{la}}
\newcommand{\pesc}{p_\text{esc}}
\newcommand{\vecrla}{\vec{r}_\text{la}}
\newcommand{\vecrc}{\vec{r}_\text{c}}
\newcommand{\nint}{N_\text{int}}

\begin{document}

   \title{Unbiased Monte Carlo continuum radiative transfer in optically thick regions}
        
        \titlerunning{Monte Carlo radiative transfer in optically thick regions}

   \author{A.~Krieger
          \and
          S.~Wolf
          }

   \institute{University of Kiel, Institute of Theoretical Physics and Astrophysics, Leibnizstrasse 15, 24118 Kiel, Germany, \\ \email{akrieger@astrophysik.uni-kiel.de}
}


   \date{Received 19 December 2019; accepted 11 February 2020}


  \abstract
   {
   Radiative transfer describes the propagation of electromagnetic radiation through an interacting medium. This process is often simulated by the use of the Monte Carlo method, which involves the probabilistic determination and tracking of simulated photon packages. In the regime of high optical depths, this approach encounters difficulties since a proper representation of the various physical processes can only be achieved by considering high numbers of simulated photon packages. As a consequence, the demand for computation time rises accordingly and thus practically puts a limit on the optical depth of models that can be simulated. Here we present a method that aims to solve the problem of high optical depths in dusty media, which relies solely on the use of unbiased Monte Carlo radiative transfer. For that end, we identified and precalculated repeatedly occuring and simulated processes, stored their outcome in a multidimensional cumulative distribution function, and immediately replaced the basic Monte Carlo transfer during a simulation by that outcome. During the precalculation, we generated emission spectra as well as deposited energy distributions of photon packages traveling from the center of a sphere to its rim. We carried out a performance test of the method to confirm its validity and gain a boost in computation speed by up to three orders of magnitude. We then applied the method to a simple model of a viscously heated circumstellar disk, and we discuss the necessity of finding a solution for the optical depth problem with regard to a proper temperature calculation. We find that the impact of an incorrect treatment of photon packages in highly optically thick regions extents even to optically thin regions, thus, changing the overall observational appearance of the disk. 
   
   }


   \maketitle
   
%

\section{Introduction}

   Nowadays, the most widely used method to perform radiative transfer simulations in astrophysics is the Monte Carlo approach. The area of application covers a wide range of astrophysical objects, such as protoplanetary or circumbinary disks, Bok globules, filaments, or planetary atmospheres, which are subjects of studies of various codes, for example, MC3D \citep{1999A&A...349..839W}, Mol3D \citep{Ober_2015}, and POLARIS \citep{2016A&A...593A..87R}. During these simulations, the photon packages that are generated travel through the simulated model space and interact with matter, for example dust, by scattering, absorption, and reemission. The particular path of the photon package as well as its interactions are determined in a probabilistic way. Throughout this process, thermal energy is deposited in the dust, which increases its temperature. At every point of interaction, the photon package has a chance to be absorbed or scattered and thus continue its path through the model space. Therefore, every photon package has to be tracked from its first emission to the moment it leaves the simulated space. The number of interactions it undergoes increases strongly with the optical depth of the medium, resulting in a high computational demand \citep[e.g.,][]{2016A&A...590A..55B,2017A&A...603A.114G,2018ApJ...861...80C}. This problem is usually handled by imposing a maximum number of interaction whose exceedance results in the removal of the photon package. This generally leads to an underestimation of the ambient temperature and consequently to an altered simulated observational appearance. Additionally, the estimation of its caused error is difficult at the very least. 
   
   Being able to simulate optically thick regions is of great relevance, in particular, if the radiation stems from these regions. For example in the case of protoplanetary disks, the heating of dust grains by deeply embedded protoplanets offers a great challenge to the Monte Carlo approach. These young accreting protoplanets are embedded in an optically thick circumplanetary disk or envelope, which effectively shields their radiation from distant regions as well as observers. In order to infer properties of these dense regions, a proper temperature calculation is crucial. Multiple approximative methods have been proposed and implemented in order to solve the problem of high optical depths, for example, the partial diffusion approximation \citep{2009A&A...497..155M}, the modified random walk \citep{2009A&A...497..155M,2010A&A...520A..70R}, or biasing techniques \citep{2016A&A...590A..55B}, each of which has its merits and drawbacks. 
   
   The diffusion approximation corrects the temperature of an optically thick region after the Monte Carlo simulation. It assumes that the probability for a photon package to leave this region without an interaction is zero and it obtains a temperature by solving the time-independent radiative diffusion equation using the temperature at the boundary of this region \citep{2000A&A...359..780W,2009A&A...497..155M}. By using this method, a low number of photon packages can already lead to a smooth temperature distribution, whereas otherwise, in a regular Monte Carlo simulation a low photon count results in a high level of noise. This method gives approximated temperature values and is typically used if the smoothness of the temperature distribution is required for further analysis, for example, when iteratively solving for a vertical hydrostatic density distribution of a circumstellar disk \citep{2009A&A...497..155M}. 
   
   A different method to handle the problem of high optical depths is the modified random walk. In a region with constant density and optical properties, a sphere with a certain radius with the position of the photon package marking its center can be defined. If the optical depth of the sphere is high, the photon package scatters multiple times before it leaves the sphere for the first time. The modified random walk uses the results from the work of \citet{1984JCoPh..54..508F} for a random walk procedure representing a large number of scattering, absorption, and reemission events. Thus, it is possible to immediately move the photon package to the rim of the sphere and skip the time consuming calculation of the scattering events that the photon package would undergo on its way toward the rim. This method can greatly enhance the performance of Monte Carlo simulations \citep[e.g.,][]{2009A&A...497..155M,2010A&A...520A..70R}. It was originally developed for isotropic scattering \citep{1984JCoPh..54..508F} and can be extended to nonisotropic scattering assuming that the phase function for scattering only depends on the angle between the incident and outgoing particle direction \citep{1978wpsr.book.....I}. Unfortunately, there is no way to apply this method to Mie scattering \citep{1908AnP...330..377M}, which is anisotropic and whose outcome of a scattering event depends on the polarization state of the incident photon. Additionally, the exact properties of a photon package, which is emitted at the rim of the sphere, are not defined; this includes the package's stokes vector and direction of emission, in particular.
   
   Lastly, the biasing technique can solve the problem of high optical depth by favoring high optical depths and in turn reducing the weight of the photon package by a corresponding factor \citep{2016A&A...590A..55B}. This method is very powerful since it gives the possibility of altering the pathway of a photon package such that it caters to the individual goals of a simulation. This freedom comes with the difficulty of quantifying the actual error of the result and thus should be applied with caution. If the photon package scatters multiple times, the weighting procedure has to be applied after every scattering event, which can result in particularly high weighting factors. Furthermore, this method clearly demotes certain paths that the photon package would have traveled along by using unbiased Monte Carlo radiative transfer \citep{2016A&A...590A..55B}. 
   
   In the following, we describe a new method to solve the problem of high optical depth, which is (i) Monte Carlo based, (ii) can be used during a regular Monte Carlo simulation, (iii) does not interfere with multiple other prominent optimization techniques (e.g., continuous absorption of photon packages, immediate reemission, peel-off), (iv) does not introduce an additional bias, and (v) is (theoretically) applicable to arbitrarily high optical depths. Sect. \ref{sec:monte_carlo_random_walk} describes the basic principles of a Monte Carlo radiative transfer code. It follows a derivation of the necessary variables needed for the implementation of our method as well as a proof of their scalability, which is crucial since it ensures the viability of the method. We proceed to describe the method and discuss its limitations as well as upgrades before we measure its performance in a testbed. In Sect. \ref{sec:application} we apply our method to a simple model of a viscously heated circumstellar disk. We compare the results of the basic Monte Carlo method with the results of our method and study the impact of a limit on the number of interactions on the calculated disk temperature. We then point out the necessity of using a method that is capable of dealing with high optical depths. We discuss the short as well as the long range impact of an inaccurate treatment of high optical depths on the resulting disk temperature and thus on simulated observations of state-of-the-art instruments, such as ALMA \citep{2002Msngr.107....7K}, PIONIER \citep{2011A&A...535A..67L}, and MATISSE \citep{2014Msngr.157....5L}. In Sect. \ref{sec:summary}, we summarize the results of this paper.

\section{Monte Carlo random walk}
\label{sec:monte_carlo_random_walk}
        For the numerical implementation and performance test of the new method, we used the three-dimensional Monte Carlo radiative transfer Code Mol3D \citep{Ober_2015}, which is a successor of MC3D \citep[][]{1999A&A...349..839W,Wolf_2003}. It achieves a high computational efficiency by combining a locally divergence free continuous absorption of photon packages \citep{1999A&A...344..282L} as well as immediate reemission according to a temperature corrected emission spectrum \citep{2001ApJ...554..615B}. Details about the code can be found in \citet{Ober_2015}.
        
        In this work, we describe a method that has been implemented in Mol3D, which uses precalculated results of Monte Carlo simulations of photon packages traveling through a homogeneous sphere of constant density, temperature, and optical properties in order to solve the problem of high optical depths in a dusty medium. We discuss the applicability of the method to arbitrarily high optical depths in arbitrary media as well as the feasibility in deriving an error estimate.

        The basic concept of this method is to precalculate the random walk of a photon package traveling from the center of a sphere to its rim while keeping track of essential variables that are needed during a radiative transfer simulation. Since this process is repeated multiple times during a Monte Carlo radiative transfer simulation, we can choose a randomly selected outcome from our precalculation. The difficulty with this approach is to find a meaningful and efficient way to store the data of possible outcomes for this process that are additionally quickly accessible. The most intuitive way is to store the energy $E_{\text{abs}}^{\text{tot}}$ deposited in the sphere as well as the complete list of properties of the photon package at the moment it penetrates the rim of the sphere. This would include the wavelength $\lambda$, the Stokes vector $\mathbf{S}=(I,Q,U,V)^\text{T}$, as well as the angle of penetration $\theta_{\text{out}}$. Certainly, the outcome of this simulation depends on the temperature as well as the radius and density of the sphere. These variables a priori depend on each other, thus, they cannot be stored for each variable separately. When binning the outcome and storing their emerging probability distributions, the memory demand easily becomes enormous depending on the accuracy we intend to achieve, thus, making a different approach desirable.
        
\subsection{General aspects of a Monte Carlo radiative transfer simulations}
\label{sec:basic_mc}
         During a Monte Carlo radiative transfer simulation photon packages travel through a grid composed of cells for which each has constant physical properties. After the emission of a photon package, the direction and the optical depth it travels is chosen in a probabilistic manner determining its next location of interaction. Between two events of interaction, energy is stored in every dust grain inside a traversed cell according to the absorption probability of the photon package by a single grain. This probability depends on the path length $l$ the photon package travels inside the cell. The deposited energy is given by \citep{1999A&A...344..282L}
\begin{equation}
\Delta E = E_\gamma \frac{C_{\text{abs}}l}{V_{\text{cell}}}, \label{eq:delta_e_dust_grain}
\end{equation}  
        where $C_{\text{abs}}$ is the wavelength-dependent absorption cross-section, $V_{\text{cell}}$ is the volume of the cell, and $E_\gamma$ is the energy the photon package carries. Throughout this process, the energy of the photon package is kept constant. When traveling through an optically thick cell, the photon package may interact multiple times before crossing one of the boundaries of the cell. Immediately after an absorption event, the photon package is reemitted isotropically with a randomly assigned direction and wavelength ($\lambda$) according to its ambient dust temperature ($T_i$) dependent spectrum. Under the assumption of a radiative equilibrium, the probability density function (PDF) $P(\lambda \,|\,T_i)$ for the wavelength is given by 
        \begin{equation}
        \frac{dP(\lambda\,|\,T_i)}{d\lambda} = \frac{C_{\text{abs}}}{K_1} \left(\frac{dB_\lambda}{dT} \right)_{T=T_i},
        \label{eq:bjorkman_and_wood}
        \end{equation}
        in which $B_\lambda$ is the Planck function and $K_1=\int_0^\infty C_{\text{abs}}(dB_\lambda/dT)d\lambda$ is a normalization factor \citep{2001ApJ...554..615B}. 
        For better performance during a simulation, this PDF and its corresponding cumulative distribution function (CDF) are precalculated in Mol3D for a set of wavelengths and temperatures. Despite the discrete sampling of reemission spectra, the resulting temperature of dust in a cell is a continuous function. We note that throughout the course of this paper, we refer to this general description of Monte Carlo radiative transfer as the basic method. Furthermore, since we do not make use of any of the aforementioned biasing techniques, we also refer to this method as unbiased.
        
\subsection{Precalculated random variables}
\label{sec:precalculated_random_variables}

        For our method, we precalculated the outcome of a simulation of photon packages traveling through a sphere of specific (constant) properties and used it to generate multidimensional binned CDFs, which account for the different possible outcomes of the simulation. During the actual Monte Carlo radiative transfer simulation, we then made use of the precalculated CDFs in order to quickly move photon packages through optically thick regions. The quantities we tracked during the precalculation are the total absorbed energy of the sphere ($\Delta E_{\text{abs}}^{\text{tot}}$), the location of the last event of absorption before the photon package leaves the sphere for the first time ($r_{\text{la}}$), as well as its wavelength when penetrating the rim of the sphere ($\lambda$). For a medium of constant and homogeneous optical properties, the distribution of these variables solely depends on the radius ($r_{\text{sph}}$), temperature ($T$), and number density ($n_{\text{dust}}$) of the sphere. The photon package is assumed to be unpolarized after reemission at the last location of absorption. Therefore, the distribution of polarization states is trivial and does not have to be tracked. However, over the course of our precalculations, we made use of Mie scattering including the proper treatment of the polarization states of photon packages. In the following, we introduce three random variables that are crucial for our method, the effective optical depth $\hat{\tau}_{\text{ext}}$, the minimal missing absorption optical depth $\taumis$, and the relative total absorbed energy $X$.

\subsubsection{Effective optical depth $\tauext$}
 We performed the calculations for spheres of specific optical depths ($\hat{\tau}_{\text{ext}}$) and the set of temperatures that have already been used previously to precalculate the PDF from Eq. \eqref{eq:bjorkman_and_wood}. 
        For the simulated spheres, we chose radii that correspond to optical depths of 
        \begin{equation}
        \hat{\tau}_{\text{ext}}^{\text{sph}} = \hat{C}_{\text{ext}}\,r_{\text{sph}}\,n_{\text{dust}} \in \mathcal{T}_{\text{sph}} \equiv \left\{ 10^{i}\, \mid \, i\in \left\{1,1.5,2,2.5,3\right\}\right\},
        \label{eq:tau_ext_definition}
        \end{equation}
        where $n_\text{dust}$ is the dust number density and $\hat{C}_{\text{ext}}$ is a temperature-dependent effective extinction cross-section. We define the latter by
        \begin{equation}
        \hat{C}_{\text{ext}}(T_i) = \frac1{K_2}\int_0^\infty C_{\text{ext}}\left(\frac{dB_\lambda}{dT} \right)_{T=T_i} d\lambda,\label{eq:effective_cext}
        \end{equation}
        where $C_{\text{ext}}$ is the wavelength-dependent extinction cross-section and $K_2=\int_0^\infty (dB_\lambda/dT)d\lambda$ is a normalization factor. 
        
\begin{figure}
   \centering
   \includegraphics{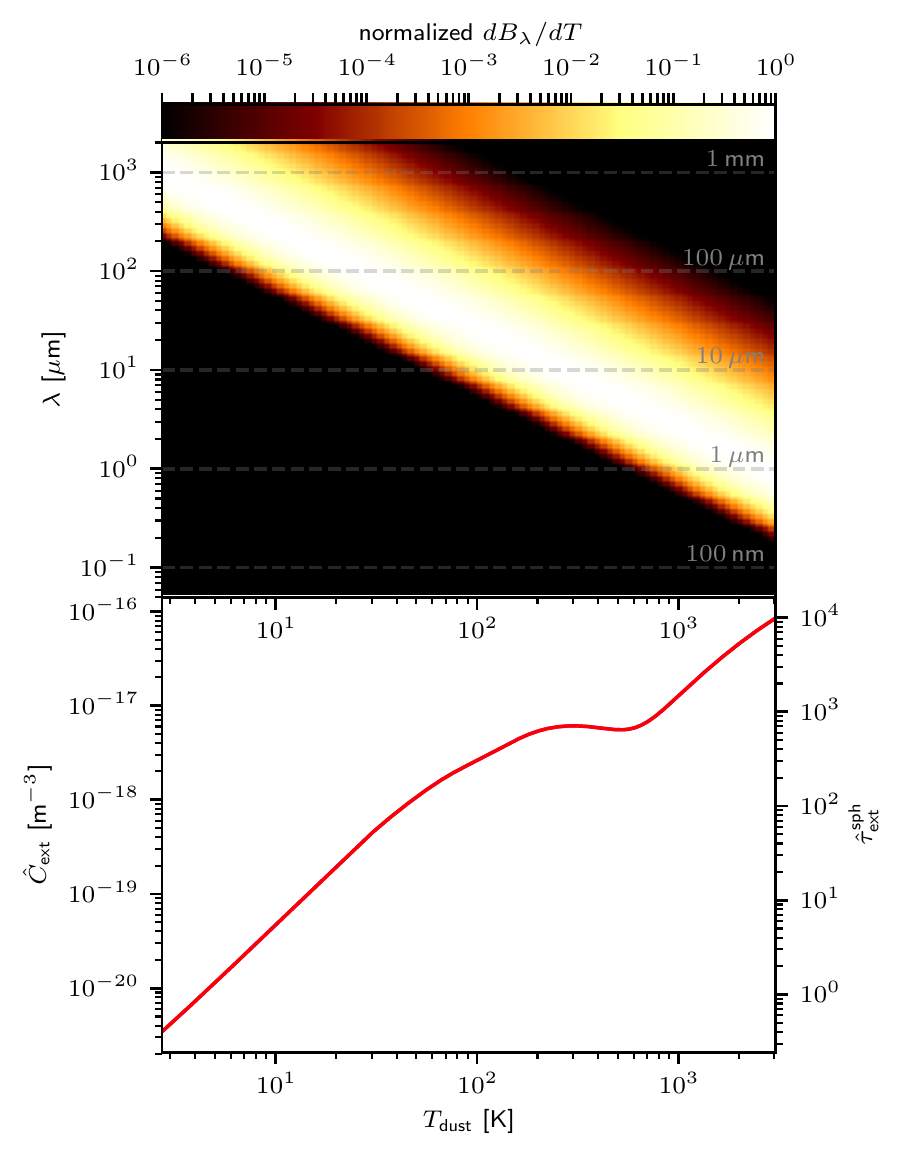}
   \caption{Visualization of the temperature dependence of reemitted photon packages when using the (difference) spectrum $dB_\lambda/dT$. \textit{Top:} The difference spectrum $dB_\lambda/dT$ that is normalized for each temperature value $T_\text{dust}$ to a maximum value of 1 is displayed. \textit{Bottom:} The effective extinction cross-section $\hat{C}_{\text{ext}}$ (see Eq. \eqref{eq:effective_cext}; left axis) and the optical depth $\tauext$ (see Eq. \eqref{eq:tau_ext_definition}; right axis) as a function of the dust temperature $T_\text{dust}$. For details, see Sect. \ref{sec:precalculated_random_variables}.}
              \label{fig:bw_spectrum_and_mean_c_ext}%
   \end{figure}

        For illustrative purposes and to stress the meaning of the optical depth $\hat{\tau}_{\text{ext}}^{\text{sph}}$, Fig. \ref{fig:bw_spectrum_and_mean_c_ext} displays properties of reemitted photon packages and their experienced optical properties. The top plot displays the difference spectrum $dB_\lambda/dT$ that is normalized to a maximum value of 1 for each temperature value $T_\text{dust}$. It shows that at higher temperatures, the peak of the spectrum moves toward shorter wavelengths. In particular, the maximum of this spectrum follows a law that is similar to Wien's displacement law given by $\lambda_\text{max}\,T \simeq 2410\,\mu\text{m}\,$K, see Sect. \ref{sec:app:displacement_law} for a short derivation. Compared to Wien's displacement law, the maximum of the difference spectrum is red-shifted, which is a necessary condition for the temperature correction technique for which it is used. Additionally, in the bottom plot, we show the temperature dependence of the effective extinction cross-section $\hat{C}_{\text{ext}}$ defined in Eq. \eqref{eq:effective_cext}. Starting from $T_\text{dust}=2.73\,$K, the cross-section $\hat{C}_{\text{ext}}$ increases strongly with the dust temperature, except for a local maximum with a subsequent dip, which is a consequence of a silicate feature\footnote{For details on the optical properties, see section Sect. \ref{sec:model}.}. Since $\hat{C}_{\text{ext}}$ is proportional to the optical depth $\tauext$, see Eq. \eqref{eq:tau_ext_definition}, both quantities share the same temperature dependence. To illustrate the corresponding value of $\tauext$ in this plot (right axis), we chose the arbitrary values $\ndust=7.8\times 10^7\,\text{m}^{-3}$ and $r_{\text{sph}}=1\,$au for the dust number density and the sphere radius, respectively. Altogether, both plots show that as the dust temperature rises in a radiative transfer simulation, the probability of reemitting photon packages with shorter wavelengths increases and, as a consequence, the effective optical depth, that is, the opaqueness of the medium, which is experienced by these photon packages, increases as well. Thus, in order to achieve a certain level of opaqueness at lower temperatures, the number density needs to be much higher compared to the case of high temperatures. Overall, the reason for using this temperature-dependent extinction optical depth $\tauext$ rather than an optical depth for a specific wavelength is that it is a much better measure for the spheres optical thickness with regard to the radiation that it produces at its particular temperature, that is, similar values of $\hat{\tau}_{\text{ext}}^{\text{sph}}$ at different temperatures correspond to a similar computational effort. 
        
        We used $501$ logarithmically sampled temperatures $T$ ranging from $2.7\,$K to $3000\,$K and $132$ wavelengths $\lambda$ between $50\,$nm to $2\,$mm and kept the radius of the sphere fixed at the arbitrary value $r_{\text{sph}}=1\,$au. In Sect. \ref{sec:scalability} we show the scalability of our simulations with regard to the sphere radius and dust density. 
        
\subsubsection{Minimal missing absorption optical depth $\taumis$} The radius of the last event of absorption $r_{\text{la}}$ strongly depends on the optical depth of the sphere. Higher values of $\hat{\tau}_{\text{ext}}$ push the last location of absorption closer toward the rim of the sphere. We chose the minimal missing absorption optical depth $\tau_{\text{abs}}^{\text{mis}}$ that the photon package has to overcome in order to leave the sphere as a measure for $r_{\text{la}}$. It is given by
        \begin{equation}
        \tau_{\text{abs}}^{\text{mis}} = \left(r_{\text{sph}}-r_{\text{la}} \right)C_{\text{abs}} n_{\text{dust}}.
        \label{eq:tau_abs_mis_definition}
        \end{equation}  
We note that $C_{\text{abs}}$ depends on the wavelength $\lambda$ the photon package has when escaping the sphere, which also corresponds to the wavelength it acquires at its last event of reemission. The only means of interaction between these two events is scattering, which leaves the wavelength of the photon package unaffected. We used 100 logarithmically sampled bins with a maximum value of $\tau_{\text{abs}}^{\text{mis}}=30$. We note that the use of an analogously defined missing extinction optical depth $\tau^\text{mis}_\text{ext}$ instead of the missing absorption optical depth $\taumis$ is conceivable; however, $\taumis$ shows to be a reliable measure and we do not expect any advantages in using $\tau^\text{mis}_\text{ext}$.

   \begin{figure*}
   \centering
   \includegraphics{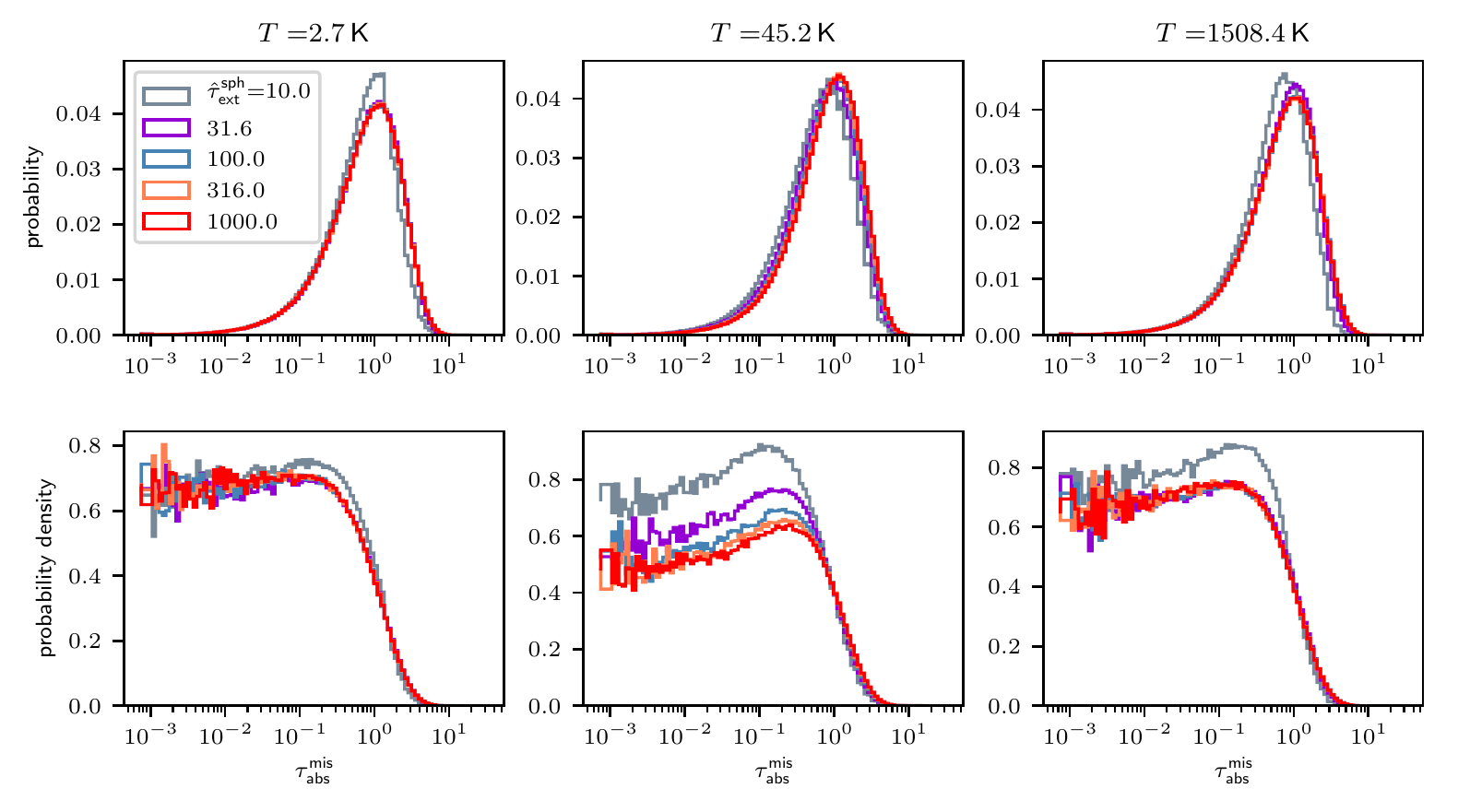}
   \caption{Result for the distribution $\taumis$ using $10^6$ simulations per temperature and a value of $\tauext$. Three different temperatures $2.73\,$K, $45\,$K, and $1508\,$K are displayed in the first, second, and third columns, respectively; the probability is shown in the first row and the probability density is illustrated in the second row. Different sphere sizes $\tauext$ exhibit notable similarities with regard to the shape and average value of their corresponding $\taumis$ distributions. This shows that radiation, which is emitted by dense spheres of different sizes but with identical temperatures, thus, originates from a similar distance from the rim of each sphere.}
              \label{fig:tau_abs_mis}%
   \end{figure*}        
        
In Fig. \ref{fig:tau_abs_mis} the probability distributions (upper row) for three different temperatures $2.73\,$K, $45\,$K, and $1508\,$K and different values of $\tauext$ are presented. The underlying probability distributions for $\tau_{\text{abs}}^{\text{mis}}$ show, in accordance with our expectation, striking similarities among all temperatures and spheres of sufficiently high values of $\tauext$. In the case of $\tauext=10$, the resulting distributions differ the most. This effect stems from the wavelength-dependent optical properties of the dust. The last location of absorption can only be as deep as the sphere's center, resulting in a strict upper limit for $\taumis$, which falls below $\taumis=10$ in certain wavelength regimes and thus causes the differences seen in Fig. \ref{fig:tau_abs_mis}. After performing ${\sim}10^{11}$ simulations for different sphere sizes and temperatures, we find that not a single photon package originated from an optical depth of $\tau_{\text{abs}}^{\text{mis}}\geq 30$. Since the probability distribution is a bin size dependent function, we calculated its corresponding probability density displayed in the second row of Fig. \ref{fig:tau_abs_mis}. Photon packages leaving the sphere have the highest probability of being absorbed and reemitted from an optical depth of about $\taumis=0.1$. The average optical depth, nonetheless, is close to $\taumis=1$. Due to the method of immediate reemission \citep{2001ApJ...554..615B} and the fact that the photon package starts its path at the center of a sphere, high values of $\taumis$ are favored. A photon package may only stem from a low value of $\taumis$ if all previous reemission events from further inside the sphere failed to transport it through the rim of the sphere. However, the fact that a photon package may scatter counteracts this trend as the geometrical reach of a photon package from one event of emission to a subsequent event of absorption decreases.

\subsubsection{Relative total absorbed energy $X$} The total absorbed energy of the sphere $\Delta E_{\text{abs}}^{\text{tot}}$ due to a single photon package depends on the pathway segments the photon package travels along inside the sphere and is given by
        \begin{equation}
        \Delta E_{\text{abs}}^{\text{tot}} = E_\gamma n_{\text{dust}} \sum_i C_{\text{abs}}(\lambda_i) l_i, \label{eq:tot_abs_energy}
        \end{equation}
where $\lambda_i$ denotes the wavelength of the photon package after its $i$-th event of interaction and $l_i$ is its corresponding pathway segment. 

   \begin{figure}
   \centering
   \includegraphics{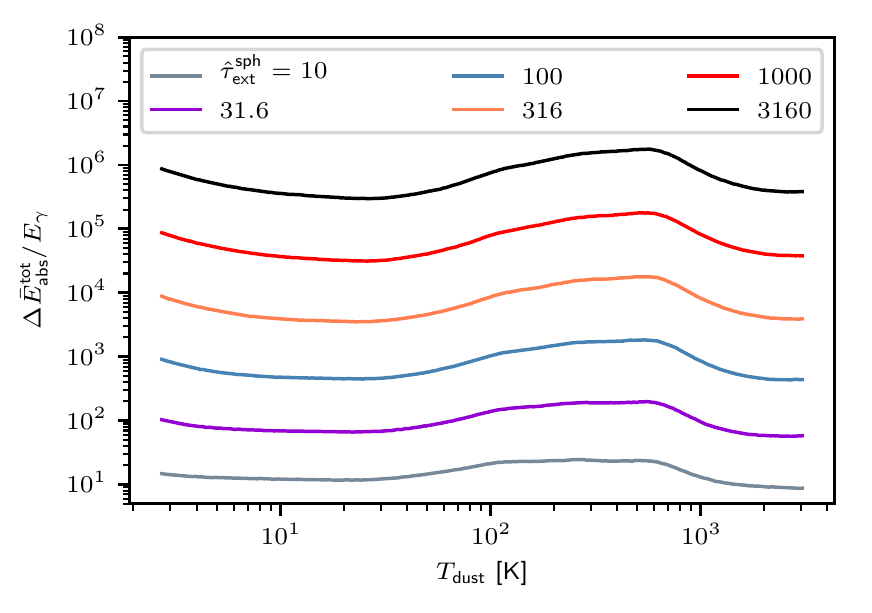}
   \caption{Average relative deposited energy using $10^4$ simulations per $\hat{\tau}^{\text{sph}}_{\text{ext}}$ and temperature value.}
              \label{fig:e_abs_tot_vs_temp}%
   \end{figure}

\begin{figure}
   \centering
   \includegraphics{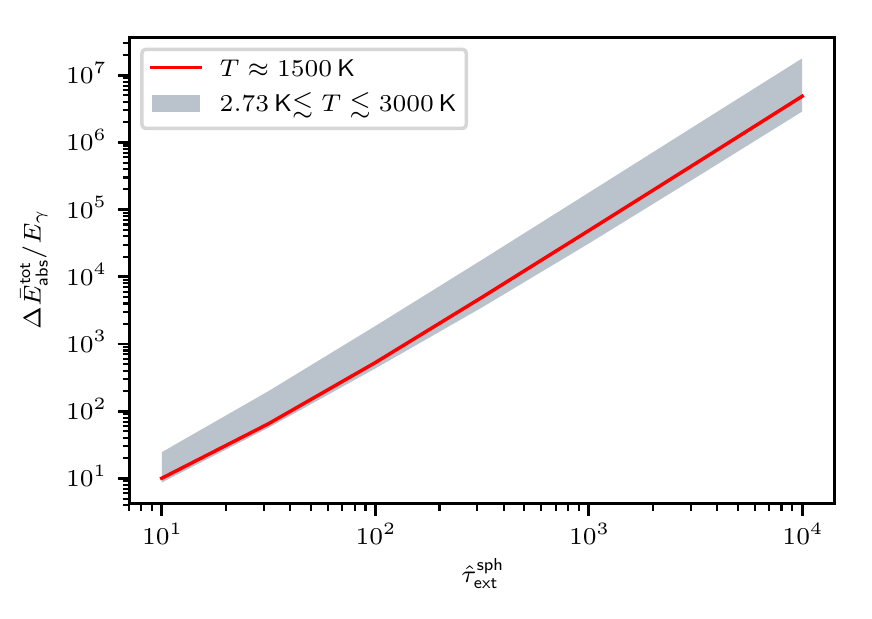}
   \caption{Extinction optical depth $\tauext$ dependence of the average total relative absorbed energy per photon package. The red line exemplarily shows the behavior for a $T\approx 1500\,$K sphere. The gray area marks the region for $\Delta \bar{E}_{\text{abs}}^{\text{tot}}/E_\gamma$ found for the complete range of temperatures between $2.73$ and $3000\,$K.}
              \label{fig:e_abs_tot_vs_tau_ext}%
   \end{figure}

Figure \ref{fig:e_abs_tot_vs_temp} shows the average relative energy a photon package deposits in a sphere for different temperatures and sphere sizes. All depicted sphere sizes feature a similar behavior, especially toward higher values of $\tauext$. Furthermore, due to the temperature-dependent definition of $\tauext$, each curve is located at a distinct level. Denser spheres, that is spheres with a higher value of $\tauext$, generally result in higher absorbed energies and show a nearly constant shift in the log-log plot. In Fig. \ref{fig:e_abs_tot_vs_tau_ext} we can see that the gray area, marking the region for possible values of $\Delta \bar{E}_{\text{abs}}^{\text{tot}}/E_\gamma$, exhibits a clear dependence of $\hat{\tau}^{\text{sph}}_{\text{ext}}$, in particular, 
        \begin{equation}
        \Delta \bar{E}_{\text{abs}}^{\text{tot}} \sim \left(\hat{\tau}_{\text{ext}}^{\text{sph}}\right)^2 \quad \Longrightarrow \quad \Delta \bar{E}_{\text{abs}} \sim n_{\text{dust}}r_{\text{sph}}^2.
        \label{eq:e_abs_tot_relation_tau_ext_sph}
        \end{equation} \label{secpoint:e_abs_tot_relation_tau_ext_sph}
        The latter relation in Eq. \eqref{eq:e_abs_tot_relation_tau_ext_sph} describes the average energy a single dust grain in the sphere acquires through a single escaping photon package. In order to mitigate the problem of binning potentially arbitrary high values of $\Delta E_{\text{abs}}^{\text{tot}}$, we instead binned the quantity $X,$ which is defined as follows:
        \begin{equation}
        X \equiv \frac{\Delta E_{\text{abs}}^{\text{tot}}}{E_\gamma \left(\hat{\tau}_{\text{ext}}^{\text{sph}}\right)^2} = \frac{n_{\text{dust}}}{\left(\hat{\tau}_{\text{ext}}^{\text{sph}}\right)^2} \sum_i C_{\text{abs}}(\lambda_i) l_i .
        \label{eq:x_variable_definition}
        \end{equation}
        This particular definition of the quantity $X$ serves a few purposes: (i) It is a measure for the total deposited energy in a sphere, (ii) it is independent of the photon energy $E_\gamma$ (for a comparison, see  Eq. \eqref{eq:tot_abs_energy}), and (iii) its factor of $1/\hat{\tau}_{\text{ext}}^{\text{sph}}{}^2$ results in a rather weak dependence of the sphere size (for a comparison, see  Eq. \eqref{eq:e_abs_tot_relation_tau_ext_sph}). 
        
\begin{figure*}
   \centering
   \includegraphics{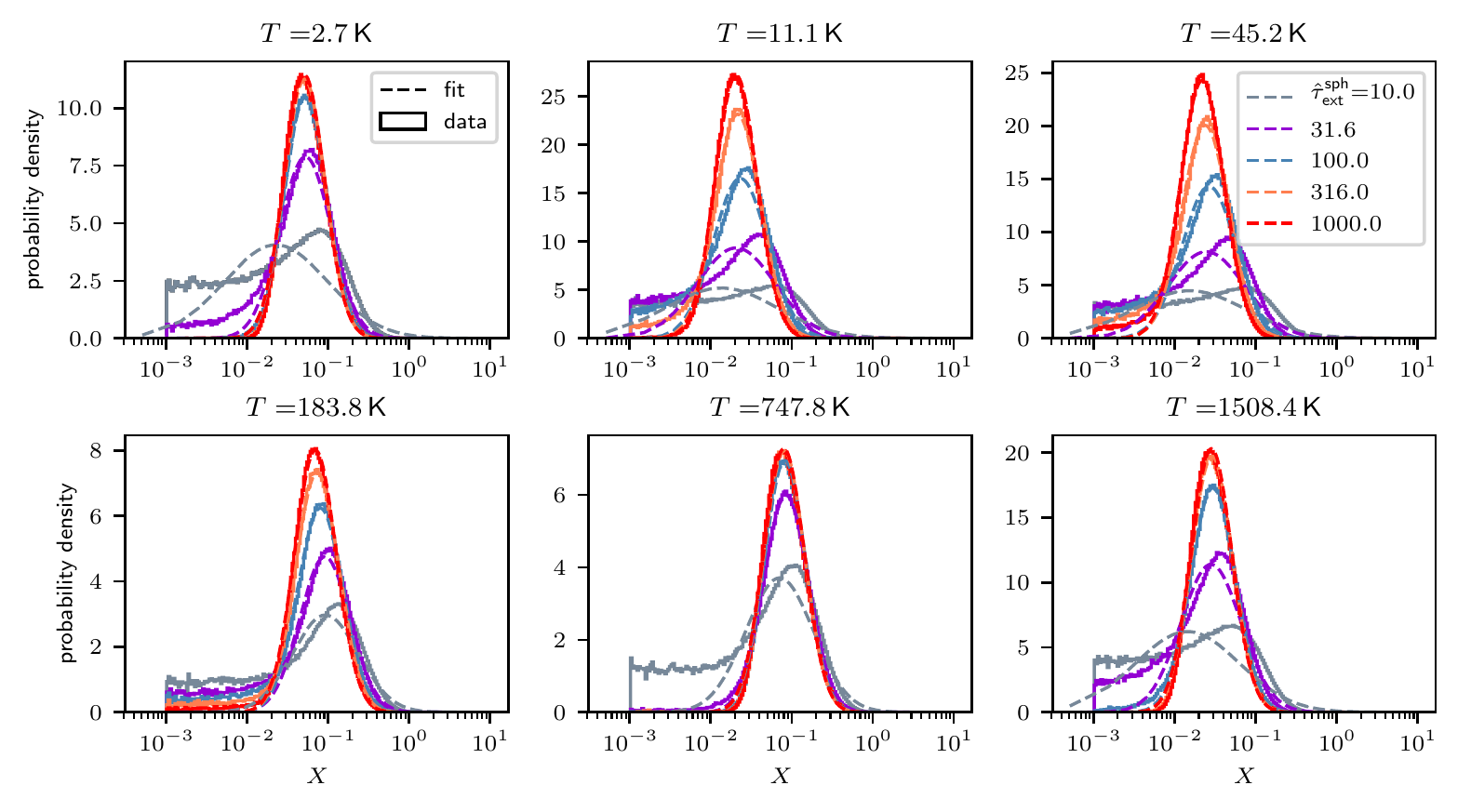}
   \caption{Results for the distributions of the variable $X$ using $10^6$ simulations per temperature and  $\tauext$ value. Outcomes for six different temperatures are displayed, which correspond to the resulting distributions (histograms) as well as their log-normal distribution fits (dashed lines) using corresponding colors. For details, see Sect. \ref{sec:precalculated_random_variables}.}
              \label{fig:lognorm_distr}%
   \end{figure*}

        In Fig. \ref{fig:lognorm_distr} we show the result of performing $10^6$ simulations per temperature and sphere size for the normalized distribution of the quantity $X$. We used $400$ logarithmically sampled bins for $X$ ranging from $10^{-7}$ to $10$ with an additional bin for $X=0$. We find that in all cases, the probability density of $X$ peaks strongly at around $X\approx 10^{-2}\ldots10^{-1}$ and only in the case of $\hat{\tau}^{\text{sph}}_{\text{ext}}=10$, the gray curve, does the distribution exhibit significant contributions from particularly low values of $X$. In addition to that, Fig. \ref{fig:lognorm_distr} shows a log-normal fit that we performed. We find that the distribution of $X$ is well described by a log-normal distribution for all cases of high optical depth. Additionally, toward higher values of $\tauext$, the distributions show indications of convergence for each individual temperature. A remarkable common property among all distributions is the lack of a $X>10$ contribution, that is, for all preformed simulations we find
        \begin{equation}
        X < X^{\text{max}} = 10. \label{eq:x_max}
        \end{equation}
        
        Overall, per calculated sphere size $\tauext$, the PDF and the CDF each have a total of $501\times 400\times 100\times 132$ ($\doteq M_T\times M_X \times M_{\tau_{\text{abs}}^{\text{mis}}}\times M_\lambda$ where $M_x$ is the number of bins for a random variable $x$) entries that have to be stored and quickly accessed during an actual Monte Carlo radiative transfer simulation. Using single precision (4 byte per entry), this corresponds to ${\sim}$10 gigabytes per sphere size $\tauext$. In realistic simulations, the dust density of cells differs and thus also the spatial extent of spheres with equal values of $\tauext$ varies on a cell basis. To avoid costly precalculations for each individual cell, which would lead to arbitrarily high memory requirements, we made use of the scalability of our simulations, which is the topic of the following subsection. Additionally, we performed a series of further tests to make sure that the number of bins, their logarithmic scaling, and full range result in acceptable coverage of the specifics of all distributions of $X$ and $\taumis$. We find that this specific sampling (i) gives reliable results and good coverage for all distributions while (ii) also keeping the memory requirements reasonably low. In particular, firstly, using a linear sampling instead increases the error of the distributions averages, secondly, a higher number of bins does not justify the resulting memory consumption, and thirdly, a lower number of bins leads to a a higher error of the average value of distributions of both random variables $X$ and $\taumis$.

\subsection{Scalability}        
        \label{sec:scalability}
        In a Monte Carlo radiative transfer simulation, the path a photon package travels in a homogeneous environment, with nonzero density, is determined by and thus equivalent to a sequence of random numbers. The optical depth $\tau$ a photon package travels is a random variable that solely depends on one of these random numbers. During our precalculations, we kept the optical properties of the environment fixed. Therefore, the path length $l$ that a photon package travels only depends on the random variable $\tau$ and the ambient dust number density $\ndust$ according to $l\sim \tau/\ndust$. Since it does not appear elsewhere in the path determination, the number density of dust only effects the path by scaling its length. Thus, given a realization of the aforementioned sequence of random numbers, the derived quantity $l\cdot\ndust$ takes the same value independent of $\ndust$. Consequently, the precalculated distributions of $\tauext$, $\taumis$, and $X$ (see Eqs. \ref{eq:tau_ext_definition}, \ref{eq:tau_abs_mis_definition}, and \ref{eq:x_variable_definition}) are also independent of $\ndust$. This is one crucial property that makes our method viable in the first place. 
        
\begin{figure*}
   \centering
   \includegraphics{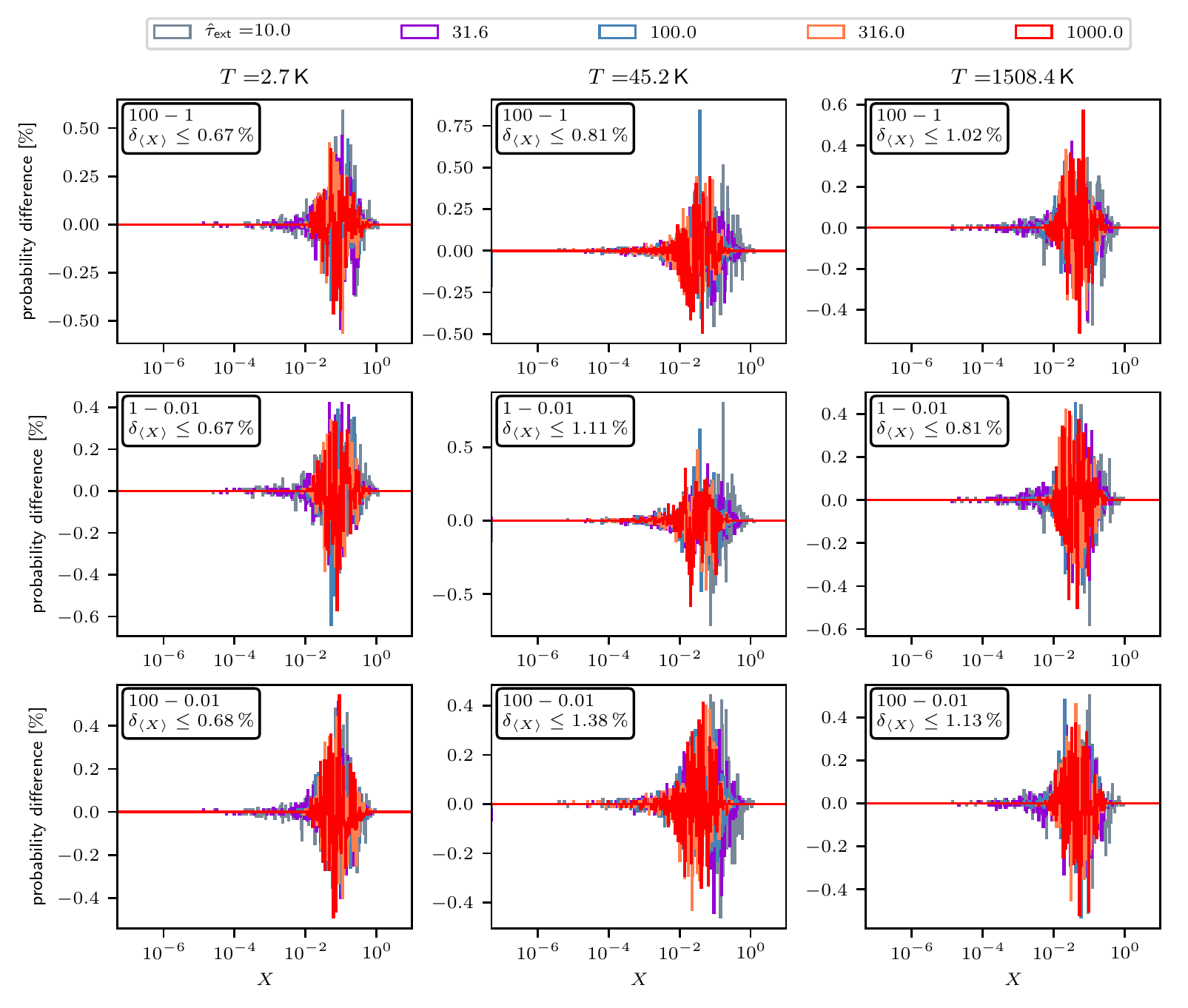}
   \caption{For three different temperatures $2.73\,$K, $45\,$K, and $1508\,$K in the first, second, and third columns, respectively, show the probability differences between distributions of $X$ using simulations with different sphere sizes $\rsph$ but equal values of $\tauext$. For every temperature and sphere size, $10^4$ simulations were performed. Text boxes contain information about the compared sphere sizes. For example, '100 - 1' denotes the comparison of the distribution using a sphere size of $\rsph=100\,$au with the distribution using a sphere size of $\rsph=1\,$au. In addition to that, the relative difference of the averages of two compared distributions is given by $\delta_{\left\langle X\right\rangle}$ in the upper left corner of each plot. Low values of $\delta_{\left\langle X\right\rangle}$ indicate a higher level of agreement.}
              \label{fig:scaling_test_1e4}%
   \end{figure*}

        In Fig. \ref{fig:scaling_test_1e4} we show results of a scalability test for the distribution of $X$. For the calculation of the distribution, we used sphere sizes of $r_{\text{sph}}=1\,$au and adjusted the density such that $\tauext$ reached the specific values from Eq. \eqref{eq:tau_ext_definition}. We then performed the same simulations, but we altered $r_{\text{sph}}$ values by a factor of 100 and 0.01 while keeping $\tauext$ fixed and readjusting the number density $\ndust$. Figure \ref{fig:scaling_test_1e4} exemplarily shows the probability differences between simulations with a different sphere size
and density for three different temperatures $2.73\,$K, $45\,$K, and $1508\,$K. The first row compares the simulations using $\rsph=100\,$au with the simulations using $\rsph=1\,$au. The second and third rows compare $\rsph=1\,$au with $\rsph=0.01\,$au and $\rsph=100\,$au with $\rsph=0.01\,$au, respectively. We find that the differences are on the order of ${\sim}1\%$. Additionally, for every subplot and value of $\tauext$, we calculated the relative differences of the average value of $X$. The maximum value of the relative difference is displayed in each subplot and is also on the order of ${\sim}1\%$. Considering the stochastic nature of this process and the fact that the number of counts per bin follows a Poisson distribution, this value equals the expected S/R of $1/\sqrt{10^4} = 1\%$. The differently scaled models are thus in agreement with our expectation. As an additional test, we increased the number of simulated photon packages by two orders of magnitude and we obtain a higher level of agreement for the distributions of one order of magnitude, which can be seen in Fig. \ref{app:fig:scaling_test_1e6} in the Appendix. This shows that the error of the precalculated distributions behaves as expected and that scalability is given.

\subsection{Method}    
        \label{sec:method_limits}
        During a Monte Carlo radiative transfer simulation, a photon package travels through a grid composed of cells with homogeneous physical properties. Succeeding an event of emission and given that the photon package is in a sufficiently large distance from the cells boundaries, we can make use of the results of our precalculations. For the highest performance of this method, we have to find the largest possible precalculated sphere that (i) fits into the cell and (ii) leaves the cell's physical conditions unchanged when the photon is transfered from the sphere's center to the last point of absorption. In order to meet the first (spatial) requirement, we calculated the minimal distance $l_\text{min}$ to the next cell border. Then the corresponding optical depth $\tausmax$ was calculated as follows:
        \begin{equation}
        \tausmax = \hat{C}_{\text{ext}}\, l_{\text{min}}\, n_{\text{dust}} .
        \end{equation}
To meet the requirements, the chosen size of the sphere has to satisfy the condition $\tauext \leq \tausmax$. The second condition requires the constancy of all physical properties that effect the path of a photon package during its way to the sphere's rim. The physical conditions inside the cell (optical properties, reemission spectrum, dust number density) are homogeneous but not necessarily constant since the cell temperature changes during the run of a simulation and as do its spectrum of reemitted photon packages and effective optical properties, see for example Eq. \ref{eq:effective_cext}. However, the effective optical properties were precalculated for each of the 501 temperature values and they only change when reaching the next temperature bin. Therefore, there is a maximum energy $\deltaemax$ that can be deposited in a single dust grain before the next temperature bin is reached. Using Eqs. \eqref{eq:delta_e_dust_grain} and \eqref{eq:x_variable_definition}, the second (temperature) condition thus requires
        \begin{equation}
        \left(\hat{\tau}_{\text{ext}}^{\text{sph}}\right)^2 \overset{!}{<} V_{\text{cell}}n_{\text{dust}}\frac{\Delta E_{\text{T}}^{\uparrow}}{E_\gamma} \frac{1}{X} .
        \end{equation}  
        In making use of the result from Eq. \eqref{eq:x_max}, that is the value of $X$ has an upper limit, this requirement implies a maximum optical depth $\tautmax$ that must not be exceeded by a chosen sphere in order to leave the cell's optical properties unaffected. The maximum optical depth $\tautmax$ is thus given by
        \begin{equation}
        \tautmax = \left( \frac{V_{\text{cell}}n_{\text{dust}}}{X^{\text{max}}}\frac{\Delta E_{\text{T}}^{\uparrow}}{E_\gamma} \right)^{1/2}.\label{eq:tau_ext_t_limit}
        \end{equation}
Therefore, for the highest efficiency possible, we chose a precalculated sphere size $\tauext$ given as follows:
        \begin{equation}
        \hat{\tau}_{\text{ext}}^{\text{sph}} = \max_{\tau \in \mathcal{T}_{\text{sph}}}\left\{\tau\,\big\vert\, \tau \leq \min(\tausmax,\tautmax) \right\}. \label{eq:tau_ext_limit}
        \end{equation}
If no such sphere has been calculated, it either means that the photon package is too close to a cell boundary or that the next temperature level is close. This implies that the method provides the greatest performance boost in the case of optically thick cells, that is, when the efficiency of the basic Monte Carlo method is low.

Given a value for $\tauext$ and the cells temperature, we used the precalculated multidimensional CDF to randomly choose values for $X$, $\taumis$, and $\lambda$. The value of $X$ was used to immediately increase the deposited energy of every dust grain in the corresponding cell. Using $\tauext$ and $\taumis$, the distance $\rla$ can be determined. Since this method was used after an event of isotropic reemission occurs, the direction in which the photon package is displaced was chosen isotropically. During the precalculation, the quantity $\rla$ marks the last point of absorption before the photon package left the sphere. Therefore, we had to launch the photon package at that position and transport it through the sphere's rim without another absorption on its way. In a real simulation, this is done by basic Monte Carlo transfer. This process has to be restarted until the photon package leaves the sphere without being absorbed. Only after a photon package is successful can the transfer of it continue. If the optical density of the cell is high, this method can be used multiple times to displace the photon package before eventually leaving the cell. The most time-consuming part of the method is its hit-or-miss nature. Even though it is already launched close to the rim, it can take many attempts before the photon package manages to leave it. Nonetheless, every time the method can be used prevents having to calculate a potentially high number of interactions and thereby computation time. If it were possible to remove either of the method's limitations, including grid induced or temperature induced limits, it would greatly benefit the effectiveness of our method. The grid induced restrictions are difficult to overcome without the loss of spatial resolution of the simulation. The temperature induced limitation, though, may potentially be removed by adjusting the reemission spectrum (see Eq. \eqref{eq:bjorkman_and_wood}) of previously absorbed photon packages during our precalculations. However, this is not part of this paper and may very well be part of a follow-up study in the future. 

\subsection{Optimization - Performance boosts}
\label{sec:upgrades}

One step of this method involves the absorption-free transfer of a photon package from its last location of absorption through the rim of the sphere. If the photon package is absorbed before leaving the sphere, it has to be reset to the previous location of reemission. Two major issues arise that significantly lower the chance of a successful escape of the photon package: (i) the optical depth and (ii) the initial direction of the photon package. The probability distribution of the distance a photon package travels before its absorption is given by an exponential function. It becomes increasingly difficult to transport a photon package over a large distance, which can lead to a high number of failed attempts. The direction of emission of the photon package is chosen randomly according to an isotropic distribution. The chance for a successful escape, however, may strongly depend on the direction of emission. Thus, a large portion of photons are initially sent in a direction with a very low escape rate. In general, it is crucial to implement optimization techniques that alleviate this problem. In fact, the viability of our method is largely grounded in their utilization. However, there are simple solutions that help to speed up this costly procedure significantly. In the following, we present three different techniques that serve this purpose.

\subsubsection{Optical depth problem - Splitting scattering and absorption length determination}
\label{sec:splitting}

A photon package with wavelength $\lambda$ is emitted and follows a path along which it may encounter several points of interaction. The type of interaction (absorption or scattering) is determined randomly by using the albedo $A=C_\text{sca}/C_\text{ext}$. Thus, if $A\neq 1$, it may happen at any point of interaction during the photons path that it is absorbed and does not escape the sphere successfully. Therefore, a large portion of photons may not even have a chance to escape the sphere as their traveled distance before absorption does not reach the minimum distance between the sphere's rim and the location of emission. In order to avoid sending out these photon packages, we split the determination of scattering and absorption events. Instead of using the extinction cross-section $C_\text{ext}$ as a measure for the interaction rate and subsequently deciding the type of interaction, we used the absorption and scattering cross-sections $C_\text{abs}$ and $C_\text{sca}$, respectively, and determined the traveled distances that lead to events of absorption and scattering independently. A short proof for the viability of this approach can be found in the Appendix in Sect. \ref{sec:app:split}.

During the simulation, the photon package was emitted inside the sphere. Its location of last absorption was determined according to its absorption optical depth $\taumis$. Subsequently, another absorption optical depth $\tau_\text{abs}$ was randomly chosen according to the absorption cross-section and dust number density. Escaping photon packages with $\tau_\text{abs}<\taumis$ can immediately be discarded. Thus, rather then limiting the probability density of $\tau_\text{abs}$ to values ${\geq}\taumis$, we instead used the sum of $\taumis$ and $\tau_\text{abs}$ as a measure for the propagated path length $l_\text{abs}$. In doing so, no photon package was created at the point of last absorption, which does not have a chance of leaving the sphere successfully. Until traversing the distance $l_\text{abs}$, the only means of interaction is scattering. If the photon package reaches $l_\text{abs}$ before leaving the sphere, it has to be reset. Additionally, whenever a photon package arrives at a scattering point and before it scatters, it is determined whether the photon package can potentially reach the sphere's rim after the forthcoming scattering event. If its remaining path length before absorption does not suffice in overcoming the remaining minimal distance to the rim of the sphere, the photon package is immediately reset to its point of emission. The importance of this step becomes apparent when considering a high optical depth of, for example, $\taumis=3$. In this case, the probability for $\tau_\text{abs}<\taumis$, which are certainly unsuccessful attempts, is ${\gtrsim}95\%$. This step of optimization is crucial for our method.

\subsubsection{Initial direction problem - Precalculated direction distribution}
\label{sec:initial_direction}

When the photon package inside the sphere is reemitted after an absorption event, the direction of emission follows an isotropic distribution. If the position of the last location of emission $\vecrla$ coincides with the center of the sphere described by $\vecrc$, the escape probability $\pesc$ for the photon package, which generally depends on the direction emission, is the same in any direction. If this is not the case, that is, if $\vecrla \neq \vecrc$, the setup is axially symmetric along the axis defined by $\vecrla-\vecrc$. Thus, the direction-dependence of $p_\text{esc}$ reduces to a dependence of the launching angle $\theta$ alone. The launching angle $\theta$ is the angle about which the direction of emission differs from the direction of the shortest distance to the rim of the sphere and is defined by the following equation:
\begin{equation}
\cos\left(\theta\right) = \frac{\left( \vecrla-\vecrc \right)\cdot \vec{v}_\gamma}{\norm{\vec{\vecrla-\vecrc}}\, \norm{\vec{v}_\gamma}},
\end{equation}
where $\vec{v}_\gamma$ is the direction of emission of the photon package. When launching a photon package close to the rim of a sphere with a relatively high value of $\taumis$, the probability may peak strongly at $\theta=0$ since shorter distances have a higher chance of being traversed without absorption. When randomly choosing $\theta$ according to an isotropic distribution, a large portion of launched photon packages therefore have low escape probabilities. This problem can be solved by precalculating the probability distribution for a successful escape for every possible sphere size, wavelength, and value $\taumis$. We used $181$ bins for the launching angle $\theta,$ ranging from $0$ (forward emission) to $\pi$ (backward emission), and set up simulations to determine the probability distribution. Compared to the size of the already precalculated CDF, which is described in Sect. \ref{sec:precalculated_random_variables}, this CDF increases the memory consumption by ${\sim}0.1\%$. 
In order to calculate the CDF properly, it is important to consider that the escape probability for every $\theta$-bin has to be weighted with the emission probability of a photon in this bin. In the case of isotropic emission, the PDF for $\cos(\theta)$ is constant.

During the radiative transfer simulation, the above distribution is used whenever the photon package is emitted at its last point of absorption $\rla$. It is important to keep the randomly chosen value of $\theta$ fixed during all escape attempts. In order to avoid performance drops, we restarted this process with newly assigned random values when a high number of attempts was exceeded. Since this number was barely reached, we were not able to find a measurable impact of this additional parameter on the resulting temperature as well as on the emission properties of the sphere. Nontheless, the distributions that are obtained from precalculated spheres are completely generated without it.

\subsubsection{Infinite beam splitting}
\label{sec:mean_of_x}

Whenever our method is used and a photon package is transported from the center of the sphere $\vecrc$ to its last location of absorption $\vecrla$, a value for $X$ is randomy chosen. By doing so, a single randomly chosen value for the amount of deposited energy is determined and increases the deposited energy of the corresponding cell. Instead of choosing a random value for $X$, it is possible to choose the mean $\left\langle X\right\rangle$ instead. This corresponds to a scenario in which the photon package is split at the center of the sphere into an infinite number of photon packages, which all follow their individual paths through the sphere until they eventually meet and combine at the location of last absorption. Since we assume that Monte Carlo radiative transfer simulations converge to the proper solution when increasing the number of photon packages, this approach decreases the level of noise of a Monte Carlo simulation. While the computation speed is not necessarily boosted significantly, the memory requirements for the results of our precalculations decrease strongly, in our case, by a factor of $400$.

\subsection{Performance test}
  \label{sec:performance}
  In this section, we showcase the performance of this method and compare it to the basic Monte Carlo method, see Sect. \ref{sec:basic_mc}, before we eventually apply it to the exemplary case of a circumstellar disk in Sect. \ref{sec:application}. The testbed that we use consists of a dense isothermal sphere with a radius of $\rsph=1\,$au and a constant density corresponding to an effective extinction optical depth of $\tauext=10^{3.5}$ at a temperature of $T=1500\,$K. The bottom plot of Fig. \ref{fig:bw_spectrum_and_mean_c_ext} shows the resulting temperature dependence of $\tauext$. We chose this particular setup for the following reasons. Firstly, the value of $\tauext=10^{3.5}$ exceeds the optical depth of our precalculated spheres (see Eq. \eqref{eq:tau_ext_definition}). Secondly, in a realistic simulation, the precalculated sphere sizes ideally cover the full range of potentially encountered optical densities. Thus, this value of $\tauext$ only exceeds the largest precalculated sphere size by half an order of magnitude. 
  
\begin{figure*}
   \centering
   \includegraphics{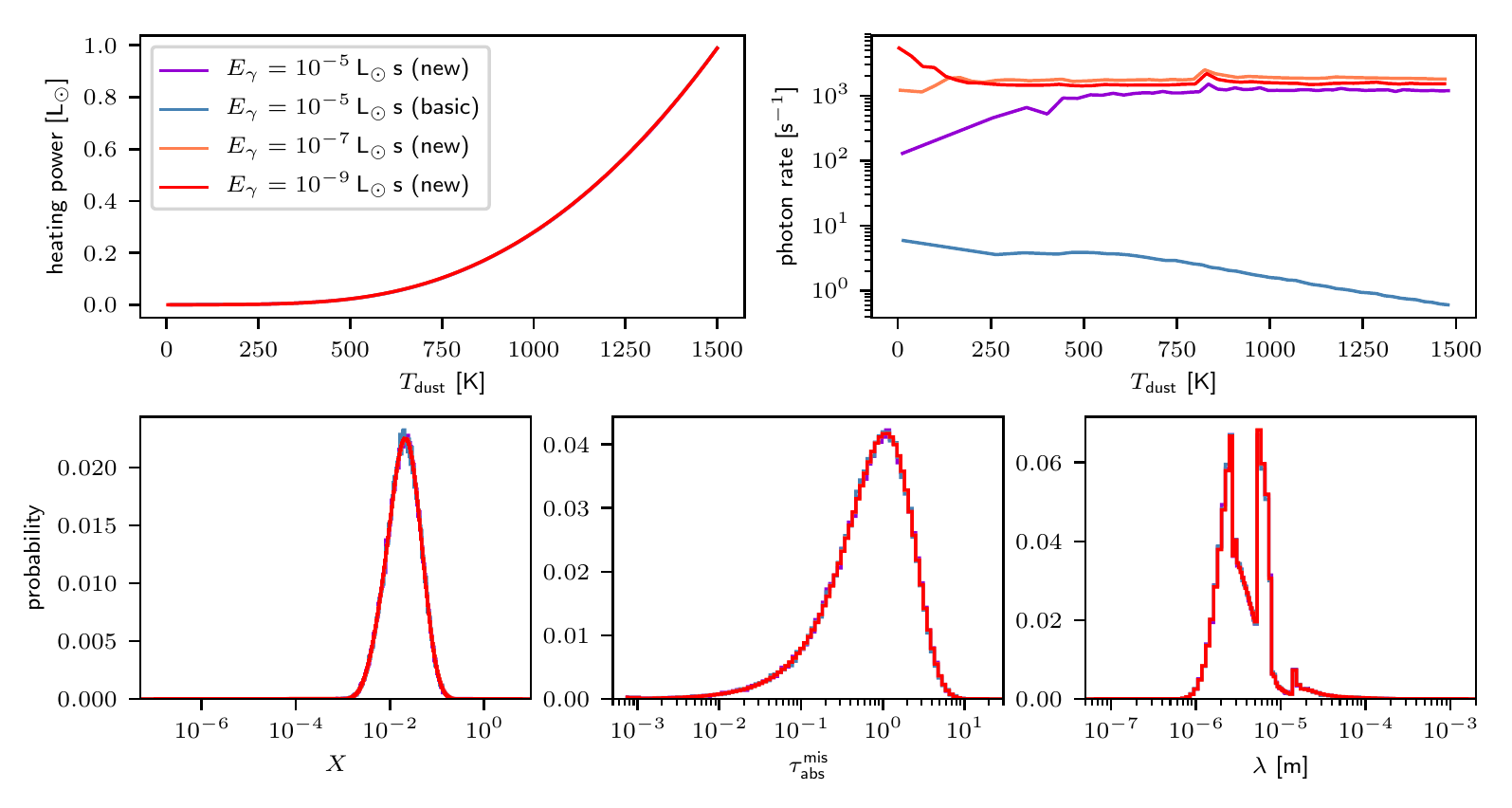}
   \caption{Viability and performance test -- comparison between the basic Monte Carlo method and our method. \textit{Setup:} A dense sphere with a radial effective extinction optical depth $\tauext(T=1500\,\text{K})=10^{3.5}$ is heated by a central source from $2.73\,$K to $1500\,$K. \textit{Upper left:} Heating curves of the sphere show high level of agreement.\, \textit{Upper right:} Performance comparison.\, \textit{Bottom:} Emitted photon statistic shows high level of agreement.}
              \label{fig:perfomance_test}%
   \end{figure*}

  The initial temperature of the dense sphere is $2.73\,$K. Throughout the simulation, photon packages were emitted isotropically at the center of the sphere. The initial wavelengths were randomly chosen according to the latest temperature during the simulation run. Launched photon packages move inside the dense sphere until they penetrate its rim. Since the temperature of the sphere is low at the start of the simulation, the spectrum of emitted photons initially peaks at a relatively long wavelength and thus these photon packages encounter a rather low dust extinction cross-section (for a comparison, see  Fig. \ref{fig:bw_spectrum_and_mean_c_ext}). While the temperature increases with the number of photon packages that were created, the dense sphere becomes increasingly opaque for the newly created photon packages on average. Eventually, it reaches its maximum effective optical depth at a temperature of $1500\,$K. During the whole process, we tracked the properties of escaping photon packages in terms of $X$, $\taumis$, and $\lambda$ as we did with the precalculated spheres. Additionally, we tracked the number of photon packages needed to heat up the dense sphere from its initial temperature to any temperature up to $1500\,$K. The photon package energy $E_\gamma$ together with the tracked number of photon packages corresponds to a central heating luminosity, which is required to sustain the corresponding sphere temperature. 
  This simulation was performed using the basic Monte Carlo method with $E_\gamma=10^{-5}$ as well as by using our method for highly optically thick regions with $E_\gamma\in \left\{10^{-5},10^{-7},10^{-9}\right\}$, where $E_\gamma$ is given in units of $\text{L}_\odot \text{s}$. For the latter case, we made use of the performance boosts described in Sections \ref{sec:splitting} and \ref{sec:initial_direction}, involving absorption-scattering splitting and an improved initial direction, respectively. Results of our comparison are displayed in Fig. \ref{fig:perfomance_test}.

  We find that the heating curves in the upper left plot of Fig. \ref{fig:perfomance_test} greatly overlap. The required heating power to reach the final temperature among all four different simulations is in good agreement and only differs by ${<}0.1\%$. Generally, relative differences between the heating curves at arbitrary temperatures are within the expected level of inherent noise due to the use of the Monte Carlo method. The three lower plots show the distributions of the tracked quantities $X$, $\taumis$, and $\lambda$ from left to right, respectively. Distributions with higher values for $E_\gamma$ require a lower number of emitted photon packages to reach any temperature level, which in turn results in a higher level of noise. Nonetheless, it becomes apparent that for all tracked quantities, the distributions overlap, thus also showing agreement between our method and the basic Monte Carlo method. We note that the shape of these normalized distributions is strongly affected by the particular bin sizes that have been used. For the sampling of wavelengths, for example, we used a catalog which samples certain domains with higher density than others, resulting in this particular shape. These tests show agreement between the heating process as well as agreement between the spectrum of the spheres and the properties of emitted photon packages. We note that in both cases, we were able to simulate the full radiative transfer, including the changing polarization state of photon packages due to scattering. The only striking difference between both methods can be seen in the upper right plot of Fig. \ref{fig:perfomance_test}. This plot shows the number of photons that we were able to transport from the center of the dense sphere outward and through its rim per second of real computation time. Certainly, the specific numbers in this plot vary for different computational architectures; the overall trends, nonetheless, are representative. The orange curve corresponds to the basic Monte Carlo method and achieves an overall rate of emitted photon packages of ${\sim}1$ photon package per second. The three other curves correspond to simulations with our newly introduced method. We find that for these three cases, the overall rate of emitted photon packages reaches ${\sim}10^3$ photon packages per second, thus a boost of three orders of magnitude. Except for the overall boost in performance, we find that this method has the positive effect of stabilizing the number of transferred photon packages during the whole run. As the temperature gradually increases, the performance of the basic method decreases. The reason for this effect is the overall increase in effective optical thickness of the sphere with temperature and its accompanying increased number of interactions. Our method, however, simply transfers the photon package close to a precalculated sphere's rim, independent of the temperature. In other words, we transport the photon package at a consistent pace in terms of spatial units while the basic Monte Carlo method transports in terms of units of the extinction optical depth. 
  
  One interesting feature of this plot is the decreased efficiency at low temperatures for high values of $E_\gamma$. The reason for this drop is the fact that our method is limited in its choice of usable precalculated sphere sizes, in this case, by the criterion of constancy of the dense spheres optical properties. The higher the energy of the photon package, the more energy is deposited in the cell, thus, resulting in a quicker increase in temperature and change of optical properties of the sphere (for a comparison, see Eq. \eqref{eq:tau_ext_t_limit}). Since the temperature levels are sampled logarithmically, this leads to a drop of the efficiency at low temperatures. Nonetheless, the method outperforms the basic Monte Carlo approach significantly, especially toward higher temperatures when dust becomes increasingly opaque. Additionally, most of the computation time is usually spent at higher temperatures, that is when our method is at its peak performance. Another feature of these curves can be seen at about $800\,$K. Here we find a peak in performance for all three curves. This is due to the fact that when starting from this temperature, the effective extinction optical depth exceeds $\tauext\approx10^3$ and thus the previously unutilized largest precalculated sphere, $\tauext=10^3$, is used.

\section{Application}
\label{sec:application}

In this section we apply our method to the situation of the radiative transfer in a viscously heated circumstellar disk and compare the results to a simulation based on the basic Monte Carlo method. Subsequently, we present a study of the impact of $\nint$ on Monte Carlo radiative transfer simulations of systems with high optical depth. 

\subsection{Model description}
\label{sec:model}
We use the following three-dimensional parametrized circumstellar disk density distribution \citep{Shakura_1973}\\
\begin{equation}
\rho(r,z) = \frac{\Sigma (r)}{\sqrt{2\pi} h(r)}\cdot \exp\left[-\frac12 \left( \frac{z}{h(r)} \right)^2 \right], \label{eq:shakura_sunyaev}
\end{equation}
 where $r$ is the radial distance from the $z$-axis at the center and $z$ is the spatial coordinate perpendicular to the midplane. The surface density distribution $\Sigma (r)$ and density scale height $h(r)$ are given by \citep{1974MNRAS.168..603L,1998ApJ...495..385H}
\begin{equation}
\Sigma (r) = \Sigma_0 \left(\frac{r}{r_0}\right)^{\beta-\alpha}\cdot\exp\left[- \left( \frac{r}{r_0} \right)^{2+\beta-\alpha} \right]\,\,\text{and}\,\,\, h(r) = h_0\left( \frac{r}{r_0} \right)^\beta,
\end{equation}
 where $\Sigma_0$ ($h_0$) is the surface density (scale height) at the reference radius $r_0$. The dimensionless parameters $\alpha$ and $\beta$ are model parameters that shape the radial density profile and flaring of the disk, respectively. The total mass of the disk $M_\text{disk}$ is obtained by the integration of Eq. \eqref{eq:shakura_sunyaev} and is composed of gas and dust with a mass ratio of $M_\text{gas}{:}M_\text{dust}=100{:}1$. In dense regions of a circumstellar disk, viscosity may become an important source of heating. Assuming a geometrically thin, optically thick accretion disk, the dissipated energy per unit area $\dot{E}_\nu$ generated through viscosity is given by \citep{1998apsf.book.....H} 
 \begin{equation}
 \frac{\dot{E}_\nu}{2} = \frac{3 G M_* \dot{M}_*}{8\pi r^3}\left[1-\left(\frac{R_*}{r} \right)^{1/2} \right] \stackrel{R_*\ll r}{\approx} \frac{3 G M_* \dot{M}_*}{8\pi r^3},
 \label{eq:viscous_heating}
 \end{equation}
 where $R_*$, $M_*$, and $\dot{M}_*$ are the stellar radius, mass, and its accretion rate, respectively, and $G$ is the gravitational constant. In order to show the applicability of our method, we performed radiative transfer simulations of solely viscously heated disks, meaning that the radiation of the central star does not contribute. 
 
 We used a spherical grid with $300$ logarithmically sampled cells in the $r$-direction and $91$ linearly sampled cells in the $\theta$-direction as well as one cell in the $\phi$-direction. The disk has an inner radius $r_\text{in}$, which is defined by the sublimation of dust at its sublimation temperature $T_\text{subl}$. The outer radius $r_\text{out}$ was chosen such that the disk extents to optically thin regions that are far from the star where the density has dropped significantly. 
 
 In our simulation, dissipated energy originates purely from the midplane cell layer and is determined by integrating equation \eqref{eq:viscous_heating} over the intersection of every cell with the midplane. For most model parameters, we chose commonly observed values that are typical of circumstellar disks around T Tauri stars \citep{2015A&A...580A..26G,2009ApJ...700.1502A}, for the specific values see Tab. \ref{tbl:parameters1}. To generate a model with efficient viscous heating, we chose a rather high value for the dust mass $M_\text{dust}$. This model results in an overall dissipated luminosity of ${\sim}1.58\cdot 10^{-3}\,$L$_\odot$. In our model, the sources of radiation are dust grains, whose spectrum is temperature-dependent. We assume spherical dust grains composed of $62.5\,\%$ silicate and $37.5\,\%$ graphite with a ratio of $\text{parallel}{:}\text{ortho}=1{:}2$ \citep{1993ApJ...414..632D} and use their corresponding mixed optical properties \citep{1984ApJ...285...89D,1993ApJ...402..441L,2001ApJ...548..296W}. Dust grains have a bulk density of $\rho_\text{bulk}=2.5\,\text{g}\,\text{cm}^{-3}$ and radii $a$, ranging from $5\,$nm to $250\,$nm following a grain size distribution given by $dn \sim a^{-3.5} da$ \citep{1977ApJ...217..425M}. Additionally, we make use of mean optical properties of dust grains \citep{Wolf_2003}. The process for scattering is described by the Mie theory. 
 
 Finally, two important parameters that limit the accuracy and have a significant effect on the computation time are the number of photon packages $N_\gamma$ and the maximum number of interactions $\nint$. The first parameter determines the energy of a single photon package $E_\gamma = L_*/N_\gamma\,\Delta t$ for the stellar radiation. Increasing the number of photon packages $N_\gamma$ leads to a lower level of noise in the temperature distribution of the circumstellar disk, but this comes with an increase in computation time. The parameter $\nint$ limits the number of interactions a single photon package can encounter before being removed from the model space. In the best case, this number is set to infinity, thus, having no undesired impact on the simulation. We note that the code was designed to ensure energy conservation and that, in the case of a viscously heated disk, every dissipating cell emits an integer number of photon packages. Consequently, the energy per photon package depends on the emission rate of the cell from which it originates and is chosen such that it is closest to the previously mentioned value, that is, $E_\gamma \lesssim L_*/N_\gamma\,\Delta t$. The way we treat radiation that is produced through viscosity is by first determining the required dust temperature of each cell in order to provide its specific luminosity. We use that temperature as an initial cell temperature. During the run of a simulation, emitted photon packages heat up the disk. During the simulation, photon packages are created according to the gradually increasing latest temperature of their corresponding cells.

\begin{table}
\begin{center}
\begin{tabular}{lccclcc}
\toprule
\multicolumn{2}{c}{Parameter} & Value && \multicolumn{2}{c}{Parameter} & Value\\
\cmidrule{1-3}  \cmidrule{5-7} 
$M_*$   &       $\left[\text{M}_\odot\phantom{\rlap{/\text{yr}}}\right]$                                 &       0.5                     && $M_\text{dust}$ &$\left[\text{M}_\odot\phantom{\rlap{/\text{yr}}}\right]$       & $10^{-3}$\\
$L_*$   &       $\left[\text{L}_\odot \phantom{\rlap{/\text{yr}}}\right]$       &       0.7                     && $r_\text{in}$&  $\left[\text{au}\phantom{\rlap{/\text{yr}}}\right]$     & $0.07$\\
$\dot{M}_*$&    $\left[\text{M}_\odot/\text{yr}\right]$                                                                 &       $10^{-9}$       && $r_\text{out}$&  $\left[\text{au}\phantom{\rlap{/\text{yr}}}\right]$    & $300$\\
\multicolumn{1}{l}{$\alpha$} &                                                                                                          &       1.8                     && $r_\text{0}$&  $\left[\text{au}\phantom{\rlap{/\text{yr}}}\right]$      & $100$\\
\multicolumn{1}{l}{$\beta$} &                                                                                                           &       1.1                     && $h_\text{0}$&  $\left[\text{au}\phantom{\rlap{/\text{yr}}}\right]$      & $7$\\
\bottomrule
\end{tabular}
\caption{Model parameters for the viscously heated disk. For details, see Sect. \ref{sec:model}.}
\label{tbl:parameters1}
\end{center}
\end{table}

 \subsection{Results}
  
     \begin{figure*}
   \centering
   \includegraphics{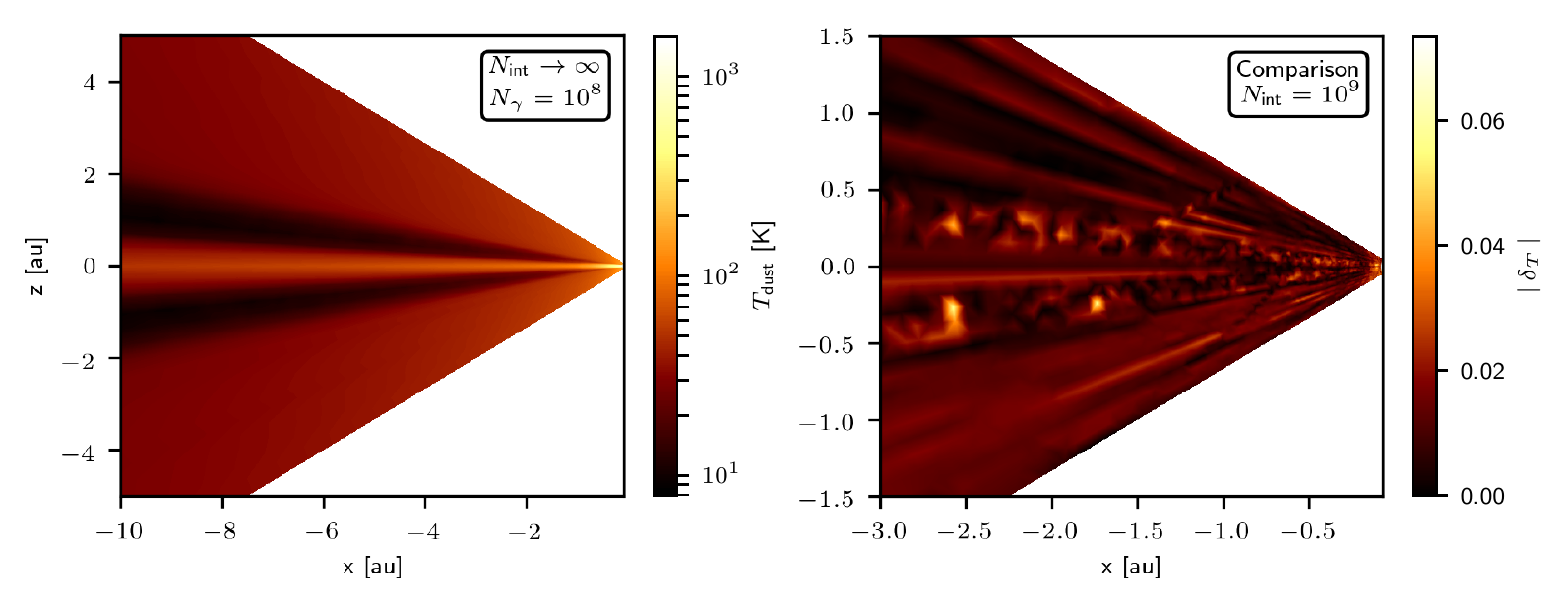}
   \caption{Comparison of the resulting temperature distributions of a viscously heated circumstellar disk. \textit{Left}:  Temperature distribution for $N_\gamma=10^8$ and $\nint \rightarrow \infty$ using our method. \textit{Right}: Relative difference between the basic Monte Carlo method and our method using $\nint = 10^9$.}
              \label{fig:viscous_disk}%
   \end{figure*}

        We performed simulations of the viscous circumstellar disk introduced in Sect. \ref{sec:model} using our method, while imposing no restriction on the maximum number of interactions, that is, $\nint{\rightarrow}\infty$ and $N_\gamma=10^8$ photon packages. The left plot of Fig. \ref{fig:viscous_disk} shows the resulting vertical cut of the temperature distribution through the midplane. In this case, within the first $0.1\,$au, multiple cells reach temperatures above the sublimation limit. As a consequence, the dust sublimates and the inner cavity size increases. As can be seen, the midplane is heated stronger than regions in the disk at higher scale heights. Looking at even higher scale heights, however, the circumstellar disk becomes optically thin and is heated by the edge of the disk's innermost region, which, overall, results in the specific temperature distribution seen in Fig. \ref{fig:viscous_disk}. This setup is symmetric with regard to the midplane with differences that arise, inherently, due to the stochastic nature of the Monte Carlo method.

        Since the basic Monte Carlo method relies on $\nint{<}\infty$ for many problems of high optical density and in order to ensure comparability, we limit our method by using the same value of $\nint{=}10^9$ as well. To do that, we have to use an estimator for the number of interactions undergone by a photon package whenever we make use of our method, that is, when transferring a photon package to the rim of a precalculated sphere. This number is not tracked during the precalculation since it increases the required memory consumption. Furthermore, its only purpose is to make sure that the number of interactions of a photon package does not exceed the limiting value $\nint$. In Sect. \ref{sec:app:estimator}, we derive an estimator for the number of interactions that we use for the comparison between both methods. The results of these simulations are shown in the right plot of Fig. \ref{fig:viscous_disk}. The plotted quantity $\abs{\delta_T}$ is the relative difference between the temperature distribution obtained by using our method $T_\text{MC+}$ and by using the basic Monte Carlo method $T_\text{MC}$. It is defined by
        \begin{equation}
        \delta_T = \frac{T_\text{MC+}-T_\text{MC}}{T_\text{MC+}+T_\text{MC}}.
        \end{equation}
   We find that the difference between both methods is on the order of ${\sim}2\,\%$ for the largest part of the plot, which is represented by dark red colors. The highest differences of ${\sim}5\,\%$, corresponding to bright yellow regions in the plot, are only found in single cells and can be explained by a low photon count in these cells. We thus find that the results of both methods are in agreement with each other. For this specific setup, we achieved a boost in computation speed by about one order of magnitude. 
   
   \subsection{Impact of imposing a maximum interaction number limit}
   \label{sec:impact_n_int}

   \begin{figure*}
   \centering
   \includegraphics{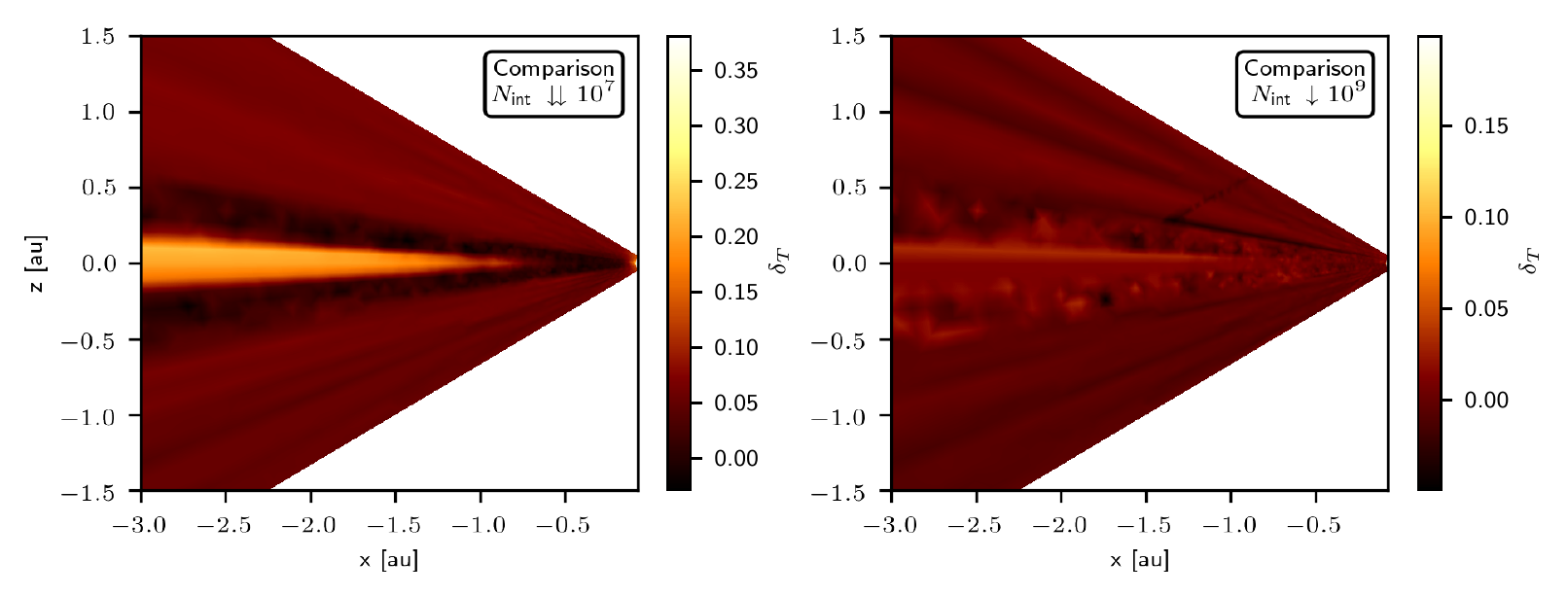}
   \caption{Studying the impact of $\nint$. Comparison of temperature distributions of viscously heated circumstellar disks. The reference model uses $\nint \rightarrow \infty$. All displayed results stem from simulations using our method. \textit{Left}: Comparison with a model using $\nint = 10^7$. \textit{Right}: Comparison with a model using $\nint = 10^9$.}
              \label{fig:viscous_disk_interactions}%
   \end{figure*}

   In this section we study the impact of the parameter $\nint$ on the resulting temperature distribution of a circumstellar disk, as described in Sect. \ref{sec:model}. Due to its reliability and the achieved boost in computation speed, the following simulations were performed using our newly developed method. In Fig. \ref{fig:viscous_disk_interactions} we see the results of a comparison between a reference simulation with $\nint{\rightarrow} \infty$ and $N_\gamma=10^8$ with simulations of a finite number of $\nint$. In the left (right) plot, we limited the number of interactions to $\nint=10^7$ ($\nint=10^9$) and plotted the relative difference to the reference model. Positive values of $\delta_T$ correspond to bright regions in the plots where the simulation with $\nint \rightarrow \infty$ reached higher temperatures. We find that the differences are larger for the lower limit $\nint=10^7$, which is in accordance with our expectation. If the number of interactions of a photon package reaches its maximum ($\nint$), the photon package was removed. This reduces the number of photons emitted from that cell, thus, leading to an underestimation of its radiation field and consequently its dust temperature. The effect of this reduced temperature is most prominent in the midplane, resulting in relative differences of up to ${\sim}1/3$. Regions in the plot with negative values, that is, higher resulting temperatures for models with a lowered value of $\nint$, are rare. The highest differences greatly exceed the level of noise seen in the right plot of Fig. \ref{fig:viscous_disk}. The right plot of Fig. \ref{fig:viscous_disk_interactions}, which  has a much higher limit of $\nint=10^9$, shows the greatest differences in cells that are close to the star, which are slightly off the midplane cell layer. These differences stem from the fact that a high number of photon packages are created in the innermost part of the disk and, in the case of a limited number of interactions, a part of them are removed from the model space without entering the regions off the midplane. On the contrary, in the reference model, these regions are heated up partly with photon packages that have undergone a number of interactions that exceed $\nint$. Nonetheless, it can be seen that the differences in the largest part of the circumstellar disk are compareable to the noise level (see Fig. \ref{fig:viscous_disk}) and only exceed this level in the innermost region of the disk. 
   
\begin{figure}
   \centering
   \includegraphics{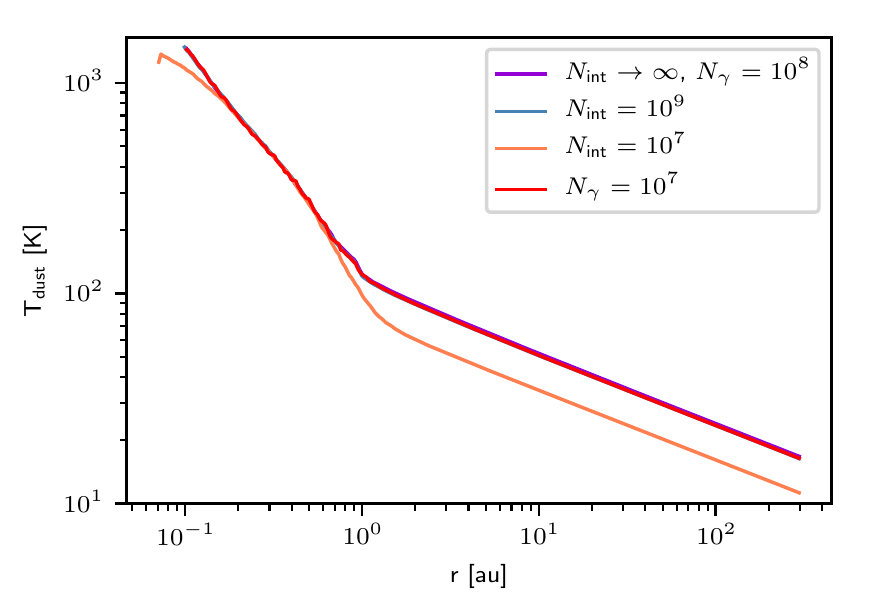}
   \caption{Impact of $\nint$ and $N_\gamma$ on the midplane temperature of a viscously heated circumstellar disk.}
              \label{fig:viscous_disk_sublimation}%
   \end{figure}

   To discuss the consequences and highlight the significance of our results, Fig. \ref{fig:viscous_disk_sublimation} shows the radial midplane temperature profile of these disks. This layer of cells produces radiation due to viscous heating. We show results for our reference model with $\nint{\rightarrow} \infty$ and $N_\gamma=10^8$ as well as for three models, each of which has one modified parameter. We find that our reference model as well as the model with a reduced number of photon packages $N_\gamma=10^7$ and the model with a reduced value of $\nint=10^9$ all produce comparable results for the midplane temperature profile. Only in the case of a further restricted value of $\nint = 10^7$ do we find strong deviations from all other simulations. We identify two major differences: (i) The innermost temperature profile and the sublimation radius are both underestimated and (ii) the temperature of the outer part of the midplane is strongly affected by the radiation originating from the inner part of the disk. The first difference occurs since the density in this part of the disk is high and photon packages may reach their interaction limit even before leaving the cell, which results in an underestimation of the temperature in this region. Consequently, a finite value of $\nint$ results in an underestimation of the inner cavity size. This effect may change simulated near-infrared long baseline interferometry observations, for instance, by MATISSE and Pioneer, and therefore impacts the interpretation of protoplanetary disk observations. The second temperature difference seen at higher radii also arises due to the removal of photon packages originating from the inner region of the disk. In the case of stellar radiation, outer regions are effectively shielded from direct irradiation. On the contrary, a viscously heated disk generates radiation throughout the extent of the disk. Thus, regions with strong viscous dissipation are shielded less efficiently form viscous dissipation than they are from stellar radiation. Due to a combination of relatively high luminosity and reduced shielding, photon packages from warm regions of the midplane reach further out laying regions and lead to a significant increase in the temperature. A limit on the number of interactions therefore underestimates the resulting midplane temperature, even at higher radii. This effect may very well be crucial, for example, in the case of simulated millimeter and submillimeter observations (e.g., by ALMA) and impact our understanding and interpretation of real observations. It is generally difficult to estimate the impact of photon packages that are removed too early and our study shows that the impact of these photon packages may very well extend to regions of a circumstellar disk, which are far away from the optically thick region from which they were emitted. 
   
   An investigation of the impact of stellar irradiation on the resulting temperature distribution shows a rather weak dependence of $\nint$, in particular, when choosing $\nint=10^7$ results in a midplane temperature profile that is already comparable to the case of $\nint\rightarrow \infty$, independent of whether the basic method or our new method was used. Furthermore, the level of noise in the optically thick regions is dominated by the inherent noise of a Monte Carlo simulation. In order to reduce the level of noise, an increased number of emitted photon packages $N_\gamma$ is required. However, the midplane temperature in our specific setup is significantly dominated by heating through viscous dissipation. Therefore, a proper treatment of optically thick regions is especially relevant when dealing with radiation that originates from deep inside these regions, such as in the case of viscous dissipation or young accreting protoplanets.

%
%

\section{Summary}
\label{sec:summary}

        In this work, we present a newly developed method to solve the problem of high optical depths in Monte Carlo radiative transfer simulations. More precisely, this method reduces the computation time with which a photon package is transported through a dense dusty medium in order to calculate the temperature distribution. We assumed homogeneously distributed dust grains and used the radiative transfer code Mol3D as a testbed for this method. We made use of precalculated results from simulations of photon packages that travel from the center of a sphere to its rim. During the precalculation, we performed Monte Carlo radiative transfer simulations to determine properties of the emitted radiation originating from a homogeneous sphere in dependence of its temperature and size. We track the relative total deposited energy in the sphere ($X$), the last location of absorption (using $\taumis$), as well as the wavelength of emitted photon packages ($\lambda$). The results are binned and stored in a multidimensional CDF. The CDF is then used to quickly transport photon packages over long distances in a probabilistic manner. We note that the precalculation was performed using basic radiative transfer, which includes wavelength-dependent optical properties, a temperature-dependent dust emission spectrum, as well as Mie-scattering with a proper treatment of the polariazation state of the photon package. Therefore, the results of these simulations are unbiased. Next, we provide a summary of our findings as well as key features of our method.
        
   \begin{enumerate}
      \item Precalculations were performed for a set of discrete values of $\tauext$, see Eq. \eqref{eq:tau_ext_definition}. Apart from the wavelength $\lambda$, we chose to track the variables $X$ and $\taumis$,  which depend on the product of the sphere radius $\rsph$ and dust density $n_\text{dust}$ and not on any one of these variables separately. In Sect. \ref{sec:scalability}, we show that the results for distributions of any value of $\tauext$, as a consequence, can be scaled and adjusted to the problem at hand, making this method viable in the first place.
      \item We find that the tracked distributions show hints for convergence for increasing $\tauext$. Interestingly, this leads to a simple relation for the mean deposited energy per dust grain and simulated photon package: $\Delta E_{\text{abs}} \sim n_{\text{dust}}r_{\text{sph}}^2$. Furthermore, we find that among all of the performed simulations, the probability for $X>10$ is very low. In fact, not even a single transported photon package among all of the performed simulations reached that value, see Sect. \ref{secpoint:e_abs_tot_relation_tau_ext_sph}. This serves as a great constraint for the maximally deposited energy per photon package, which is dependent on the temperature, density, and size of the spheres. 
      \item In Sect. \ref{sec:method_limits}, we derived and discuss the limitations of our method with regard to the change of the medium's optical properties as well as spatial limits due to a model grid, see Eq. \eqref{eq:tau_ext_limit}. Subsequently, in Sect. \ref{sec:upgrades} we also present upgrades to the method that significantly boost the effectiveness -- the splitting of scattering and absorption length (see Sect. \ref{sec:splitting}), the utilization of our prior knowledge of the initial direction (see Sect. \ref{sec:initial_direction}), as well as the infinite beam splitting (see Sect. \ref{sec:mean_of_x}).
      \item We carried out a performance test using, as a testbed, a dense sphere of optical depth $\tauext \approx 3162$ and achieve a boost of about three orders of magnitude in computational speed, see upper right plot in Fig. \ref{fig:perfomance_test}. In even denser environments, for instance, around young planets or stars, this value is easily exceeded and our method is expected to perform even better than that.
      \item Simulations for increasingly high values of $\tauext$ are time-consuming when using the basic Monte Carlo method. In the bottom plots of Fig. \ref{fig:perfomance_test}, we show that the statistic of emitted photon packages as well as the statistic of the deposited energy of the basic method and our method are in remarkable agreement. Therefore, the calculation of larger spheres can be sped up significantly when using the set of already precalculated smaller spheres. Its error has been shown to be in agreement with the variance of a Poisson distribution.  
      \item We applied our method to a simple model of an optically thick viscously heated circumstellar disk and show the validity of the method. We study the impact that the imposed maximum number of interactions $\nint$ has on the resulting temperature profile of the disk. We find that the choice of $\nint$ is decisive throughout the full extent of the disk and not only in the optically thick regions where it is generated, see Sect. \ref{sec:impact_n_int}. The impact of a value of $\nint$ that is too low is generally difficult to predict. 
   \end{enumerate}

High optical depths pose a well known problem to Monte Carlo radiative transfer simulations. This study presents a novel method that solves this problem of Monte Carlo radiative transfer simulations and thus enables us to properly predict temperatures of objects with optically thick regions.

\begin{acknowledgements}
         We thank all the members of the Astrophysics Department Kiel for helpful discussions and remarks. 
         We acknowledge the support of the DFG priority program SPP 1992 "Exploring the Diversity of Extrasolar Planets (WO 857/17-1)".
\end{acknowledgements}

\bibliographystyle{aa} 
\bibliography{literature} 

\begin{appendix}

\section{Derivation of expected optical depths}

\subsection{Interactions and absorption-scattering}
\label{sec:app:split}

The following derivation shows that the determination of absorption lengths and scattering lengths can be done independent of each other. The use of extinction lengths as well as interactions with a randomly assigned type of interaction (absorption or scattering) is equivalent to that. 

\subsubsection{Splitting absorption and scattering}
The probability density for an absorption optical depth $\tau_{\text{abs}}$ is given by $f(\tau_{\text{abs}})=\exp\left(-\tau_{\text{abs}}\right)$. The probability density for reaching a distance interval $l$ is then given by 

\begin{equation*}
p(l)=\lim_{\Delta l \rightarrow 0} \frac{P(l \in \left[ l,l+\Delta l \right])}{\Delta l} = \int_{\tilde{C}_{\text{abs}}l}^{\tilde{C}_{\text{abs}}(l+\Delta l)} f(\tau_{\text{abs}}) \,d\tau_{\text{abs}} .
\end{equation*}
By expanding $f(\tau_{\text{abs}})$ at $\tau_{\text{abs}} = \tilde{C}_{\text{abs}}l$ where $\tilde{C}_{\text{abs}}$ is the specific cross-section for absorption, we find that
\begin{equation}
p(l)=\tilde{C}_{\text{abs}}\exp\left( -\tilde{C}_{\text{abs}}l \right). \label{eq:splitting_abs_sca}
\end{equation}

\subsubsection{Using interactions}
The probability density for an interaction optical depth $\tau_{\text{ext}}$ is given by $f(\tau_{\text{ext}})=\exp\left(-\tau_{\text{ext}}\right)$. Subsequent to a reemission event, a photon package is moved along a straight line until it reaches a randomly selected extinction optical depth. At that point, another random number determines whether the subsequent interaction is an event of absorption or scattering. The probability for a scattering event is given by $\tilde{C}_{\text{sca}}/\tilde{C}_{\text{ext}}$ where $\tilde{C}_{\text{sca}}$ ($\tilde{C}_{\text{ext}}$) is the specific cross-section for scattering (extinction).

In the scattering free case, the probability density for traveling the distance $l$ after traversing the extinction optical depth $\tau_{\text{ext}}=\tilde{C}_{\text{ext}}l$ is, in analogy to the derivation of Eq. \eqref{eq:splitting_abs_sca}, given by $\tilde{C}_{\text{ext}}\exp\left(-\tilde{C}_{\text{ext}}l \right)$. For the probability density of a subsequent event of absorption, we thus find
\begin{equation}
p_0 = \tilde{C}_{\text{ext}}\exp\left(-\tilde{C}_{\text{ext}}l \right) \cdot \left(1-\frac{\tilde{C}_{\text{sca}}}{\tilde{C}_{\text{ext}}}\right) = \tilde{C}_{\text{abs}}\exp\left(-\tilde{C}_{\text{ext}}l \right).\label{eq:no_sca_eq}
\end{equation}
In order to calculate the probability for $n>0$ events of scattering before the eventual event of absorption at distance $l$, we have to integrate over the probability density of all possible paths of this kind. Let $l_i\geq 0$ denote the path length the photon package travels before its $i$-th event of scattering. It follows that after its $n$-th and final scattering event, it travels a distance of $l_a = l - \sum_{i=1}^nl_i$ before being absorbed. For the $i$-th event of scattering, the probability density is, in analogy to the derivation of Eq. \eqref{eq:no_sca_eq}, given by $\tilde{C}_{\text{sca}}\exp\left( -\tilde{C}_{\text{ext}}l_i \right)$. Due to the fact that the individual path lengths are constrained by $\sum_{i=1}^n l_i\leq l$, the probability density $p_n$ is given by

\begin{equation}
\begin{split}
p_n =& \int_{0}^{l}dl_n \tilde{C}_{\text{sca}}\exp\left( -\tilde{C}_{\text{ext}}l_n \right)\int_{0}^{l-l_n}dl_{n-1}\tilde{C}_{\text{sca}}\exp\left( -\tilde{C}_{\text{ext}}l_{n-1} \right)\cdots \\ &
 \cdots \int_{0}^{l-\sum_{i=2}^n l_i}dl_{1} \tilde{C}_{\text{sca}}\exp\left( -\tilde{C}_{\text{ext}}l_1 \right) \cdot \tilde{C}_{\text{abs}}\exp\left( -\tilde{C}_{\text{ext}}l_a \right). \label{eq:n_scattering_events}
\end{split}
\end{equation}

\noindent In integrating Eq. \eqref{eq:n_scattering_events}, one arrives at 
\begin{equation*}
p_n = \frac{l^n\tilde{C}_{\text{sca}}^n}{n!} \tilde{C}_{\text{abs}}\exp\left( -\tilde{C}_{\text{ext}}l \right).
\end{equation*}
This implies, that with the highest probability, the photon package scatters $n_\text{max}=\left\lfloor l\cdot\tilde{C}_{\text{sca}} \right\rfloor$ times. Eventually, we have to take the sum of $p_n$ over all possible numbers of scattering events and find for the probability density: 
\begin{eqnarray*}
        p(l) &=& \sum_{n=0}^\infty p_n \\
                &=& \sum_{n=0}^\infty \frac{l^n\tilde{C}_{\text{sca}}^n}{n!} \tilde{C}_{\text{abs}}\exp\left( -\tilde{C}_{\text{ext}}l \right)\\
                &=& \tilde{C}_{\text{abs}}\exp\left( -\tilde{C}_{\text{ext}}l \right)\sum_{n=0}^\infty \frac{l^n\tilde{C}_{\text{sca}}^n}{n!} \\
                &=& \tilde{C}_{\text{abs}}\exp\left( -\tilde{C}_{\text{ext}}l \right)\exp\left(\tilde{C}_{\text{sca}}l\right) \\
                &=& \tilde{C}_{\text{abs}}\exp\left(-\tilde{C}_{\text{abs}}l \right) .
\end{eqnarray*}
Since this result and the result obtained in Eq. \eqref{eq:splitting_abs_sca} are the same, we can split the calculation of absorption and scattering length. \hfill $\square$

%
%
%
%

\subsection{Estimator for $\nint$}
\label{sec:app:estimator}
        
        For comparison purposes between the basic Monte Carlo method and our newly developed method, we kept the maximum number of interactions $\nint$ equal in both simulations. In order to keep track of the number of interactions when using the new method, we estimated it every time the method was used. 
        It is possible to derive an exact estimate when assuming $\langle X \rangle \xrightarrow{\hat{\tau}_{\text{ext}}^{\text{sph}}\rightarrow \infty}X^* >0 $ for the case of $\hat{\tau}_{\text{ext}}^{\text{sph}}\rightarrow \infty$, where $\langle\cdot\rangle$ denotes the average of all possible paths of the photon package:
        \begin{equation*}
                \langle X \rangle \stackrel{\text{eq. } \eqref{eq:x_variable_definition}}{=} \left\langle \frac{n_{\text{dust}}}{\left(\hat{\tau}_{\text{ext}}^{\text{sph}}\right)^2} \sum_i C_{\text{abs}}(\lambda_i) l_i  \right\rangle = \frac{n_{\text{dust}}}{\left(\hat{\tau}_{\text{ext}}^{\text{sph}}\right)^2} \left\langle  \sum_a C_{\text{abs}}(\lambda_a) l_a  \right\rangle .
        \end{equation*}
For the latter equation, we changed the sum over interaction path lengths to a sum over absorption path lengths. This is possible since the wavelength of the photon package does not change between two consecutive absorption events. Since $\hat{\tau}_{\text{ext}}^{\text{sph}}\rightarrow \infty$ in the limit, any two consecutive absorption events are independent of each other, we can exchange the average of the sum by the sum of averages. Let $\left\langle  \cdot  \right\rangle_{\tau_{\text{abs}},\lambda}$ denote the average of all possible paths between two consecutive absorption events. It follows that
        \begin{equation*}
                \langle X \rangle = \frac{n_{\text{dust}}}{\left(\hat{\tau}_{\text{ext}}^{\text{sph}}\right)^2} \sum_a \left\langle  C_{\text{abs}} l  \right\rangle_{\tau_{\text{abs}},\lambda} = \frac{N_{\text{abs}}}{\left(\hat{\tau}_{\text{ext}}^{\text{sph}}\right)^2}  \left\langle  \tau_{\text{abs}} \right\rangle_{\tau_{\text{abs}},\lambda}.
        \end{equation*}
For the latter equation, we used $\tau_\text{abs}=\ndust C_\text{abs}l$ and replaced the sum by the total number of absorption events $N_\text{abs}$. The average is simply $\left\langle  \tau_{\text{abs}} \right\rangle_{\tau_{\text{abs}},\lambda}=1$. Therefore, we arrive at 
        \begin{equation}
                \langle X \rangle = \frac{N_{\text{abs}}}{\left(\hat{\tau}_{\text{ext}}^{\text{sph}}\right)^2}\quad \Rightarrow \quad N_{\text{abs}} =  \langle X \rangle \left(\hat{\tau}_{\text{ext}}^{\text{sph}}\right)^2.
                 \label{eq:n_abs}
        \end{equation}
        In this limiting case, the total number of interactions $\nint = N_{\text{abs}} + N_{\text{sca}}$ converges to $ \nint \rightarrow \tau_{\text{ext}}^{\text{tot}}$ since one interaction occurs per unit of $\tau_\text{ext}$ on average. Analogously to the derivation of Eq. \eqref{eq:n_abs} and using $C_\text{ext}=C_\text{abs}+C_\text{sca}$, we thus arrive at
        \begin{equation*}
        \nint = \left\langle \sum_i n_{\text{dust}} C_{\text{ext}}(\lambda_i) l_i  \right\rangle = N_{\text{abs}}\left( 1 + \left\langle  \frac{C_{\text{sca}}(\lambda)}{C_{\text{abs}}(\lambda)} \tau_{\text{abs}} \right\rangle_{\tau_{\text{abs}},\lambda} \right) .
        \end{equation*}
Since $\lambda$ and $\tau_\text{abs}$ are independent random variables, we can split the average into a product of two separate averages. Thus, 
\begin{equation*}
\left\langle  \frac{C_{\text{sca}}(\lambda)}{C_{\text{abs}}(\lambda)}\tau_{\text{abs}} \right\rangle_{\tau_{\text{abs}},\lambda} = \left\langle  \frac{C_{\text{sca}}(\lambda)}{C_{\text{abs}}(\lambda)}\right\rangle_{\lambda}, 
\end{equation*}
where $\left\langle \cdot \right\rangle_{\lambda}$ denotes the average over all possible wavelengths using the corresponding temperature-dependent PDF. With Eq. \eqref{eq:n_abs}, we find that
        \begin{equation*}
\nint = \langle X \rangle \left(\hat{\tau}_{\text{ext}}^{\text{sph}}\right)^2\left\langle  \frac{C_{\text{ext}}(\lambda)}{C_{\text{abs}}(\lambda)}\right\rangle_{\lambda}.
        \end{equation*}
Therefore, we estimate the number of undergone interactions during the transport of a photon package from the center of a sphere to its last location of absorption with
        \begin{equation*}
        \Delta \nint = \left\lceil X\cdot \left(\hat{\tau}_{\text{ext}}^{\text{sph}}\right)^2 \left\langle  \frac{C_{\text{ext}}(\lambda)}{C_{\text{abs}}(\lambda)}\right\rangle_{\lambda}\right\rceil.
        \end{equation*}

\subsection{Displacement law of the difference spectrum}
\label{sec:app:displacement_law}
In this section we study the difference spectrum $dB_\lambda/dT$ introduced by \citet{2001ApJ...554..615B}. We find that (i) $dB_\lambda/dT$ is positive for all $\lambda,T>0$, (ii) it is unimodal, that is, it has exactly one (global) maximum, and (iii) the maximum follows a law similar to Wien's displacement law, that is, $\lambda_\text{max}T=$const.
The difference spectrum is given by
\begin{equation*}
\frac{dB_\lambda}{dT} = \frac{2h^2c^3}{k_\text{B}T^2\lambda^6} e^{\frac{hc}{k_\text{B}T\lambda}} \left( {e^{\frac{hc}{k_\text{B}T\lambda}}-1} \right)^{-2},
\end{equation*}
which is strictly positive for all $\lambda,T>0$. For a fixed temperature $T$, its derivative with respect to the wavelength $\lambda$ satisfies
\begin{equation*}
\frac{d}{d\lambda}\frac{dB_\lambda}{dT} = s(T,\lambda) \left( -6x -1 + 2 \frac{e^{1/x}}{e^{1/x}-1} \right),
\end{equation*}
where $x=\frac{k_\text{B}T\lambda}{hc}$ and $s(T,\lambda)>0$ for all $\lambda,T>0$. Thus, an extremal point requires
\begin{equation*}
6x+ 1 = 2 \frac{e^{1/x}}{e^{1/x}-1}.
\end{equation*}
This equation has exactly one positive solution $x_\text{max}>0$ since (i) the left-hand side (l.h.s.) is smaller than the right-hand side (r.h.s.) as $x\rightarrow 0^+$ and (ii) the slope of the l.h.s ($m_\text{l.h.s.}$) is given by $m_\text{l.h.s.}=6$, while the slope of the r.h.s ($m_\text{r.h.s.}$) satisfies $0< m_\text{r.h.s.}<2$ for all $x>0$. Furthermore, we find that
\begin{equation*}
\frac{dB_\lambda}{dT} \stackrel{\lambda\rightarrow 0^+}{\longrightarrow} 0 \quad \wedge \quad \frac{dB_\lambda}{dT} \stackrel{\lambda\rightarrow \infty}{\longrightarrow} 0,
\end{equation*}
thus requiring the extremal point at $x_\text{max}$ to be a maximum and, consequently, $dB_\lambda/dT$ to be unimodal. A numerical calculation shows that the maximum is located at
\begin{equation*}
x_\text{max} \simeq 0.1675.  
\end{equation*}
Therefore, for a given temperature $T$,  the unique (global) maximum of the difference spectrum $dB_\lambda/dT$, which is located at the wavelength $\lambda_\text{max}$, satisfies:
\begin{equation*}
\lambda_\text{max}\,T \simeq 2410\,\mu\text{m}\,\text{K}.
\end{equation*}

        \begin{figure*}
   \centering
   \includegraphics{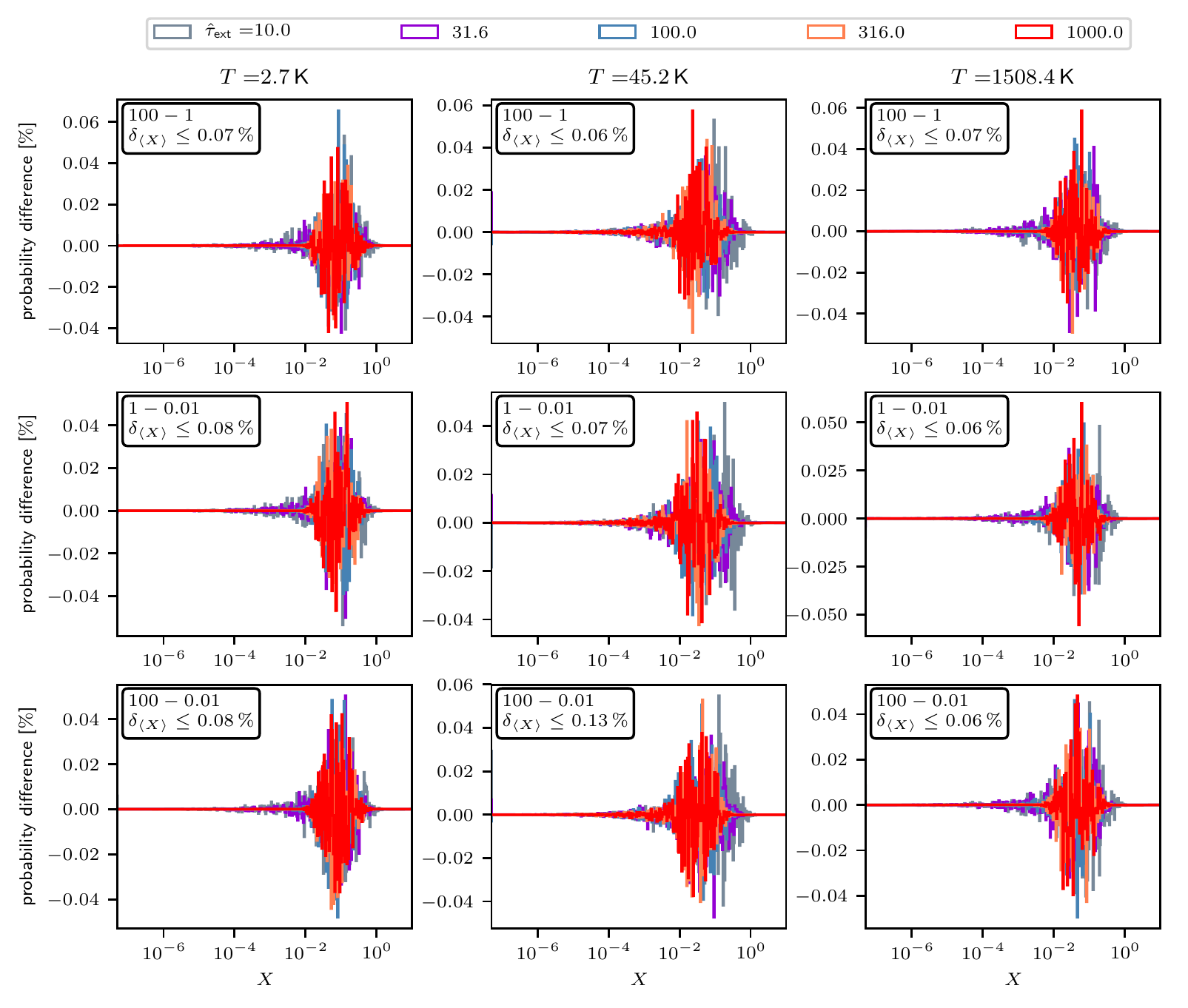}
   \caption{For three different temperatures $2.73\,$K, $45\,$K, and $1508\,$K in the first, second, and third columns, respectively, show the probability differences between distributions of $X$ using simulations with different sphere sizes $\rsph$ but equal values of $\tauext$. For every temperature and sphere size, $10^6$ simulations were performed. Text boxes contain information about the compared sphere sizes. For example, '100 - 1' denotes the comparison of the distribution using a sphere size of $\rsph=100\,$au with the distribution using a sphere size of $\rsph=1\,$au. In addition to that, the relative difference of the averages of two compared distributions is given by $\delta_{\left\langle X\right\rangle}$ in the upper left corner of each plot. Low values of $\delta_{\left\langle X\right\rangle}$ indicate a higher level of agreement.}
              \label{app:fig:scaling_test_1e6}%
   \end{figure*}

\end{appendix}

\end{document}